\documentclass[%
 reprint,
superscriptaddress,
preprintnumbers,
 amsmath,amssymb,
 aps,
prl,
]{revtex4-2}

\usepackage{ amssymb }
\usepackage{color}
\usepackage{graphicx}
\usepackage{dcolumn}
\usepackage{bm}
\usepackage{hyperref}
\usepackage[caption=false]{subfig}
\newcommand{\dt}{\widetilde{D}}

\usepackage[normalem]{ulem}

\begin{document}

\title{Entanglement Entropy Transitions with Random Tensor Networks}

\author{Ryan Levy}
\email{rlevy3@illinois.edu}
\affiliation{Institute for Condensed Matter Theory and IQUIST and NCSA Center for Artificial Intelligence Innovation and Department of Physics, University of Illinois at Urbana-Champaign, IL 61801, USA} 
\author{Bryan K.  Clark}
\email{bkclark@illinois.edu}
\affiliation{Institute for Condensed Matter Theory and IQUIST and NCSA Center for Artificial Intelligence Innovation and Department of Physics, University of Illinois at Urbana-Champaign, IL 61801, USA}
\date{\today}

\begin{abstract}
Entanglement is a key quantum phenomena and understanding transitions between phases of matter with different entanglement properties are an interesting probe of quantum mechanics.   We numerically study a model of a 2D tensor network proposed to have an entanglement entropy transition first considered by Vasseur et al.[Phys. Rev. B \textbf{100}, 134203 (2019)]. We find that by varying the bond dimension of the tensors in the network we can observe a transition between an area and volume phase with a logarithmic critical point around $D\approx 2$. We further characterize the critical behavior  measuring a critical exponent using entanglement entropy and the tripartite quantum mutual information, observe a crossover from a `nearly pure' to entangled area law phase using the the distributions of the entanglement entropy and find a cubic decay of the pairwise mutual information at the transition. We further consider the dependence of these observables for different R\'enyi entropy.  This work helps further validate and characterize random tensor networks as a paradigmatic examples of an entanglement transition.
\end{abstract}
\maketitle

\section{Introduction}
Entanglement is one of the paradigmatic examples of quantum phenomena. 
One way of quantifying the entanglement between two regions of a system is through entanglement entropy.  Quantum states can  differ fundamentally in the amount of entanglement between their two halves;  of particular interest is determining how this entanglement, and thus entanglement entropy, scales with the system size.  While a typical quantum state has entanglement which scales linearly with the size of the bipartition (i.e. volume-law scaling), ground states have entanglement which scales independent of (area-law) or logarithmically  with (log-law) the bipartition size.  Recently there has been interest in understanding phase transitions between different phases of matter which are characterized by the nature of their entanglement scaling.  Examples of such entanglement phase transitions happen in disordered quantum systems such as the many-body localized phase, random quantum circuits, and random tensor networks. 

In the many-body localized phase, one tunes disorder in the Hamiltonian finding a transition from volume law eigenstates at low disorder to area law eigenstates at high disorder \cite{mblReview}.  While these transition has been studied numerically\cite{PhysRevB.91.081103,PhysRevLett.113.107204,PhysRevB.94.184202,PhysRevLett.115.187201,PhysRevB.94.045111,PhysRevX.7.021013,PhysRevX.5.041047,PhysRevLett.119.075701,gray2019scale,villalonga2020eigenstates,PhysRevB.102.104201,villalonga2020characterizing} and via renormalization group (RG)\cite{PhysRevLett.112.217204,PhysRevX.5.031032,PhysRevX.5.031033,PhysRevB.93.224201,thiery2017microscopically,PhysRevLett.122.040601,PhysRevB.99.094205,PhysRevB.99.224205,PhysRevB.102.125134,Parameswaran2017,10.21468/SciPostPhys.10.5.107}, there are still a number of an open questions about the nature of this transition.

\begin{figure}[t!]
    \centering 
    \includegraphics[width=0.55\columnwidth]{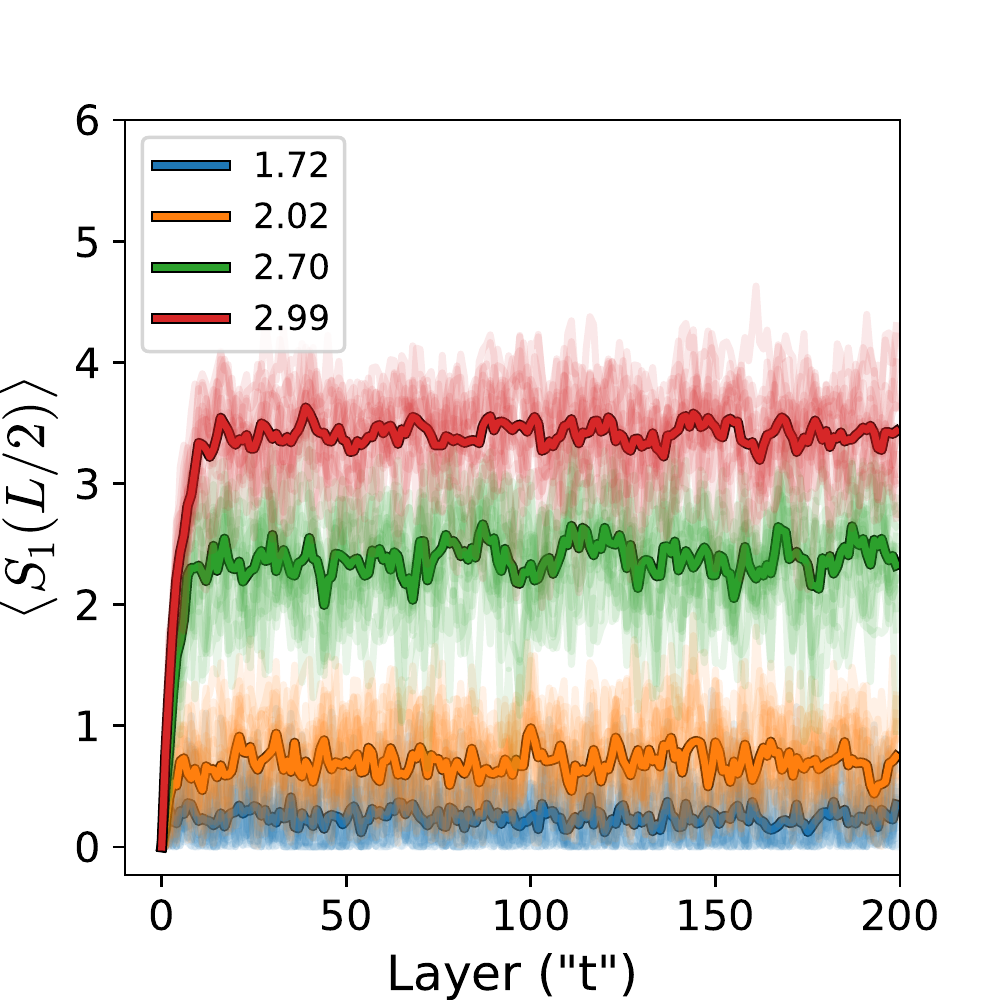}
    \includegraphics[width=0.4\columnwidth,trim=135 30 125 40, clip]{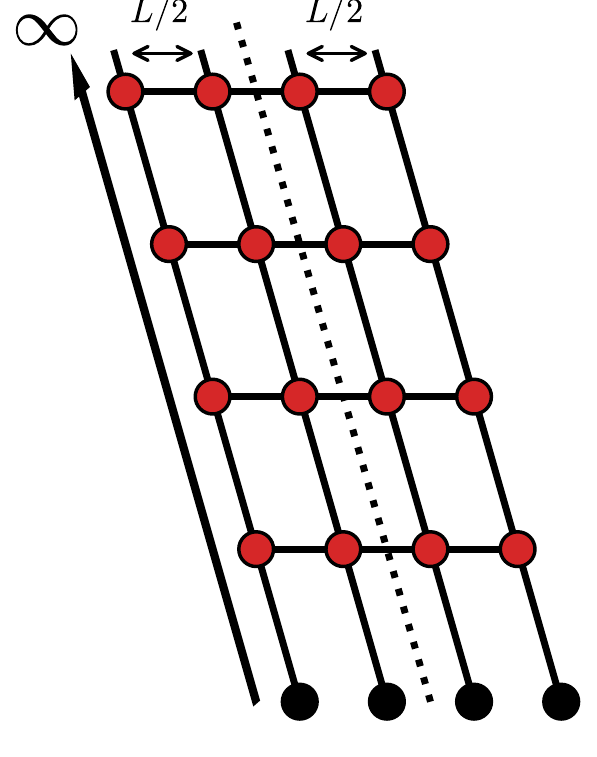}
    \includegraphics[width=\columnwidth,trim=0 80 0 50, clip]{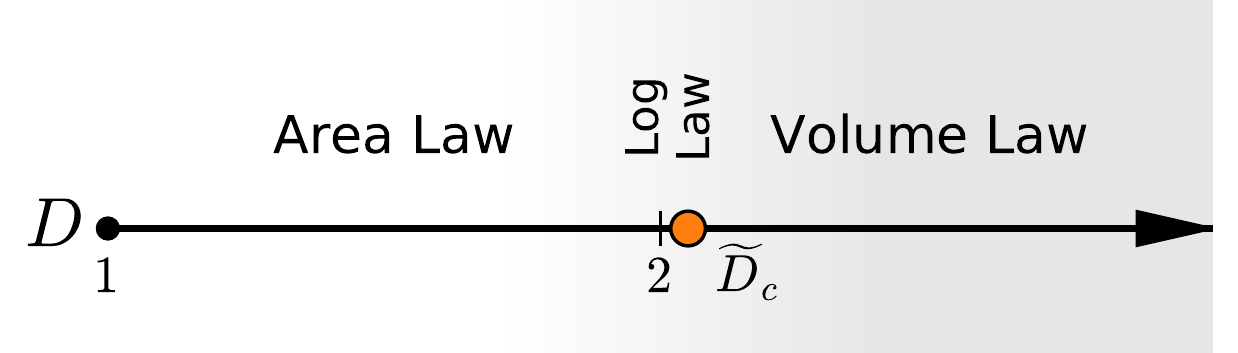}
    \caption{\textit{Top left:} von Neumann entanglement entropy at each layer of the tensor network, for 4 different $\dt$ (see legend). In bold is the average of 16 trials, each displayed lightly in the same color. 
    \textit{Top right:} cartoon of the studied tensor network. The left arrow represents the ``time'' direction, where we study the entanglement entropy layer by layer, and a dashed line represents a cut $L_A=L/2$. 
    \textit{Bottom:} Illustrated phase diagram of the random tensor network for the von Neumann entanglement entropy transition. The marker represents the critical point $\dt_c$.  Gray shading represents a non-zero leading or sub-leading logarithmic term in our fit of $\langle S_1(L/2)\rangle$. }
    \label{fig:phase}
\end{figure}

Another model also exhibiting a volume-law to area-law transition is random quantum circuits where the system is driven from volume to area-law as the fraction of local measurements is increased
\cite{PhysRevX.9.031009,PhysRevB.99.224307,PhysRevLett.125.030505,Li2019,Gullans2019a,PhysRevResearch.2.013022,Bao2019,PhysRevB.101.060301,Jian2019,PhysRevB.101.104301,PhysRevB.102.094204,PhysRevB.102.054302,ippoliti2021entanglement,PhysRevB.102.014315,PhysRevLett.125.210602,PhysRevResearch.2.043072,czischek2021simulating,block2021measurementinduced,Yang2021}.   
Between these phases in the random quantum circuit model is a critical point that has logarithmic entanglement growth. Various generalization of these quantum circuit models have also been considered \cite{PhysRevB.98.205136,10.21468/SciPostPhys.7.2.024,PhysRevB.100.064204,PhysRevB.101.235104,alberton2020trajectory,Lavasani2021,PhysRevResearch.3.023200,PRXQuantum.2.030313,minato2021fate,doggen2021generalized}.

A final model exhibiting entanglement transitions is random tensor networks.  In a random tensor network, one has a series of tensors of a given bond-dimension(s) connected in some geometry whose  values are randomly selected. The network itself is formed independent of any Hamiltonian or observable measure. Two common geometries which have been studied are tree geometries (i.e. tree tensor networks  \cite{Lopez-Piqueres2020}) and square grid geometries (i.e. PEPS \cite{Vasseur2018}).  
Interestingly, both MBL and random quantum circuits can be viewed as a transition in a PEPS random tensor networks where the Hamiltonian or quantum gates respectively are treated as tensors \cite{PhysRevB.101.235104,yu2019bulk}, where there is an extra constraint on the network, such as Hermiticity or unitarity.   

In this work, we will consider a PEPS network where the values of the tensors are normally distributed and  the (average) bond-dimension plays the role of tuning the transition.  This model was originally studied analytically in ref.~\cite{Vasseur2018} where a replica-symmetry breaking analysis argued for the existence of area to volume-law transition.  We consider this model numerically, characterizing the critical point, determining the critical bond-dimension $\dt_c$, the critical exponent $\nu$, and R\'enyi-index-dependent slope of the logarithmic critical point.  In addition, in verifying the existence of the volume-to-area law transition we verify that the $q \rightarrow 0$ replica limit from the analytical analysis is reliable. 

\section{Model}

We study a rectangular PEPS-like tensor network which consists of a tensor $T$ with normal distributed ($\mathcal{N}(0,1)$) random elements and $n$ indices $i_1,\dots,i_n$ each of bond dimension $D_j=\dim(i_j)$; for a bulk tensor $n=4$ while $n=3$ at the edge. The network is of width $L$ and $N$ layers tall with open boundary conditions; the network always begins in a product state of physical dimension $d=2$, shown in the right of fig.~\ref{fig:phase}.

A network is then further classified by the dimensions of the tensor indices. When each tensor has identical bond dimensions, $\forall j, D_j=D$, the network is denoted as a \textit{uniform} network. When not uniform, the index dimension is chosen from a log-normal distribution $\text{Lognormal}(\mu,\sigma)$.
We introduce the modified average bond dimension 
\begin{align}
    \dt = \exp\left[\sum_{k=1} P(i_j=k)\cdot \log(k) \right].
\end{align}
We heuristically find that using $\dt$ better aligns non-uniform with uniform results than using $\langle D \rangle$ (see fig.~\ref{fig:dtilde_comparison} and supplementary). This can be motivated by considering the average maximum entropy for a generic cut through the network, $\langle \log(D_j) \rangle$; thus $\dt$ is a better approximation to the bound of entanglement entropy\footnote{For a uniform network, $S_1$ is bounded by $\log(D)$, which is recovered exactly since $\tilde{D}=D$ in that case}. For all studied points $\langle D \rangle \geq \dt $. 
Reported $\dt$ values are in the infinite system limit found by computing $P(i_j=k)$ exactly and rounded unless otherwise noted. 
We note that even using this $\tilde{D}$ we find that the uniform and disordered case seem to disagree by a small shift. For example, all values of $\langle S_1(L_A) \rangle$ for all $L_A$ at fixed $L$ are indistinguishable between the uniform case at $D=2,3$ and the disordered case at $\dt=2.055,3.051$ respectively suggesting a slight shift of 0.05 between them \footnote{For illustration, $\tilde{D} \approx 2.05$ is the same as $\langle D \rangle \approx 2.1$ which is still distinctly different than $D=2$. See fig.~\ref{fig:dtilde_comparison}. }.

To determine the $n$th R\'enyi entropy $S_n^\ell$ at a given layer $\ell$, we compress each previous layer into a MPS of auxiliary bond dimension $m$ using the `density matrix method' \cite{itensor}. During calculation of the MPS we discard all singular values with weights smaller than $10^{-16}$, and unless otherwise noted truncate the bond dimension at $m=2048$.
After forming the compressed MPS, we cut the system into two subsystems of size $L_A$ and $L_B$ with $L_A\leq L_B$, measuring the entropy $S_n^\ell(L_A;L)$ for all possible $L_A$ which ranges from $1$ to $L/2$.  The R\'enyi entanglement entropy is defined by $S_n(L_A) = (1-n)^{-1}\mathrm{Tr}[\rho^n_A]$  where $n\in\mathbb{R}^+$ and $\rho_A$ is the reduced density matrix for the first $L_A$ sites in the system, i.e. the region $A$. A special case is the the $n=1$ R\'enyi entropy, or the von Neumann entropy $S_1(L_A) = -\mathrm{Tr}[\rho_A \log \rho_A ]$. The entropy values are all non-negative and monotonic for $n>1$ such that $n_2\geq n_1>1 \implies S_{n_2}(L_A) \leq S_{n_1}(L_A)$ \cite{RenyiProperties}. We average this quantity over all layers (after removing the first 50 layers) i.e. $\langle S_n(L_A)\rangle = \langle S^\ell_n(L_A;L)\rangle_\ell $, averaged again over multiple trials. We compute error bars by using trial-averaged data. An example of the entropy at each layer is shown in the top left of fig.~\ref{fig:phase}. Calculations were all performed using the ITensor library \cite{itensor}. 

This model has been studied by Vasseur et al. \cite{Vasseur2018}. Using an analytic replica approach they find a R\'enyi index $n$ dependent area to volume law transition in $\langle S_n\rangle$;  in the volume law phase at large $D$ they find that $\langle S_n \rangle \sim L_A \log D$ at large $D$.  They argue for a  logarithmic scaling critical point $\langle S_n \rangle \sim \alpha_n \log L_A+C_n +f_n(L_A/\xi) $ where $f_n$ is a universal function and $\xi \sim |D-D_c|^{-\nu}$ for some critical bond dimension $D_c$, at which $\alpha_n=1/3$ for $n\geq 1$ \cite{Jian2019}.

\section{Entanglement Transitions}
Using the method outlined in the previous section, we study the average entanglement entropy $\langle S_n(L_A) \rangle $ as well as its distribution.

\subsection{Uniform Tensor Networks}
\begin{figure}
    \centering
    \includegraphics[width=\columnwidth]{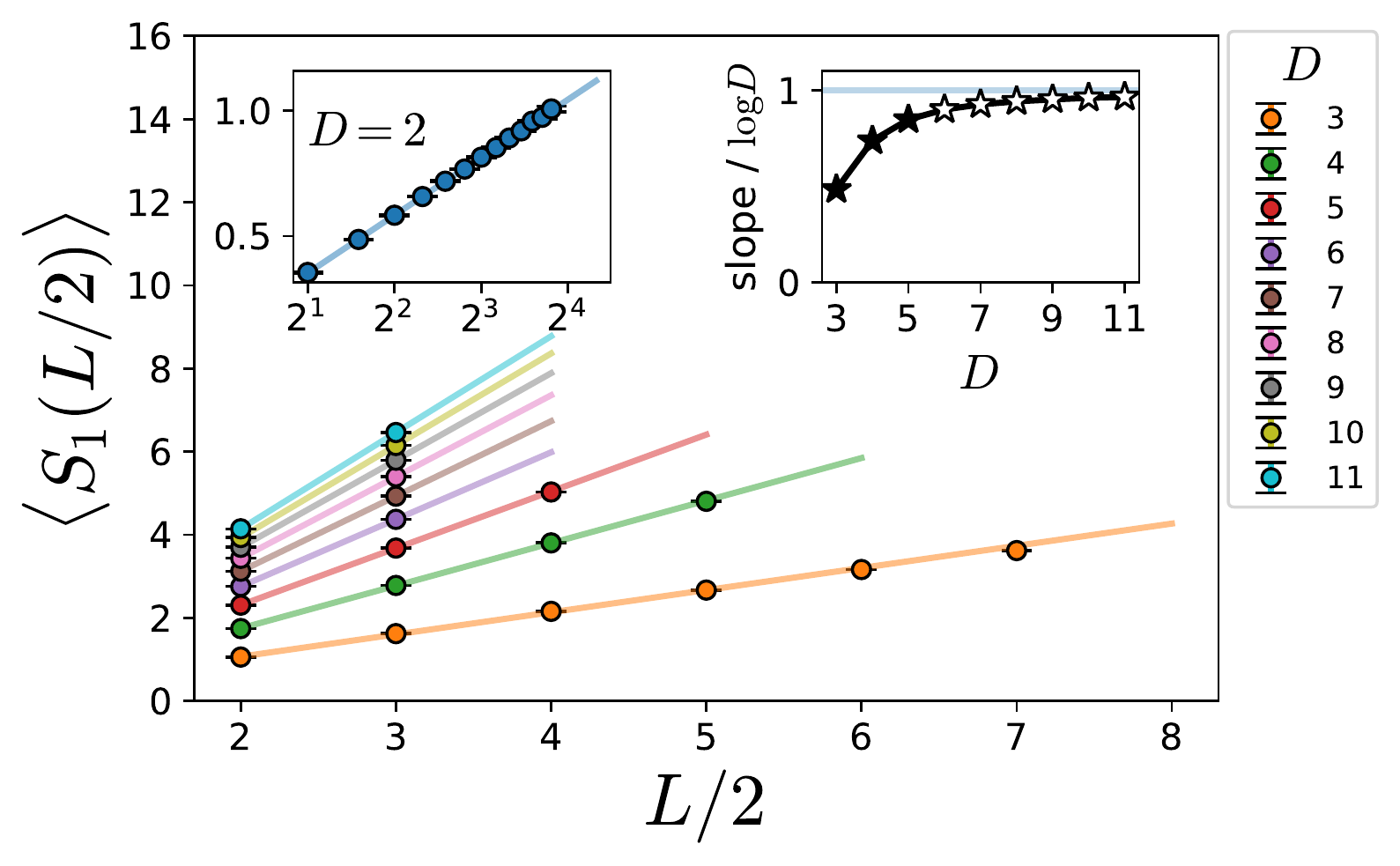}\label{fig:S_vs_L_uniform}
    \caption{Average von Neumann entanglement entropy $\langle S_1(L/2)\rangle $ for a uniform tensor network along with linear fits to the data. \textit{Left inset:} $D=2$ with a log fit.
    \textit{Right inset:} Slope of the linear fits as a ratio of the expected volume law slope $\log(D)$. Open stars indicate only two points fit to a line. }
    \label{fig:S1_uniform}
\end{figure}

We start by considering a uniform bond dimension $D$ along all legs.  At $D=1$, the network has  trivially zero entanglement.  At $D=2$, we find that the entanglement $S_1(L/2)$ scales logarithmically as  $S_1 \propto \alpha_1 \log L$ with $\alpha_1= 0.335(2)$ out to $L=28$; note that the slope of the logarithm agrees well with the predicted value at $n=1$.  We find that as $n$ increases, $\alpha_n(D=2)$ drifts from 1/3 going as $\alpha_n = 0.216(1+1/n)-0.099$ (see fig.~\ref{fig:alpha_function}), a form that is consistent with the conformal field theory ansatz discussed in ref.~\cite{PhysRevB.101.060301}.  At larger $D>2$, $S_1$ grows linearly (with potential subleading corrections) approaching the anticipated $S_1 \propto \log(D)/2 L$ at large $D$, shown in the inset of fig.~\ref{fig:S1_uniform}.  For $D>6$, the computational difficulty prevents us from measuring $L\gg1$ and at these $D$ we estimate the slope using only two points. Despite this, these two points approach the expected slope of $\log D$.

\subsection{Non-Uniform Tensor Networks}\label{sec:non_uniform}

In order to explore the behavior of the entanglement entropy around $D=2$ we turn to a non-uniform network, with tensor bond dimensions chosen from a log-normal distribution $\mathrm{Lognormal}(\mu,\sigma=0.2)$ where we tune $\mu$ to obtain different $\dt$. 

We can also provide a lower bound to the area law transition. At $\langle D \rangle = 1.5$ the bonds of $D=1$ on the square lattice leads to percolation and thus a trivial area law phase. We therefore restrict our studies above this average bond dimension.

 \subsubsection{Entanglement Entropy Statistics}
\begin{figure}
    \centering
    \subfloat{\includegraphics[width=0.5\columnwidth]{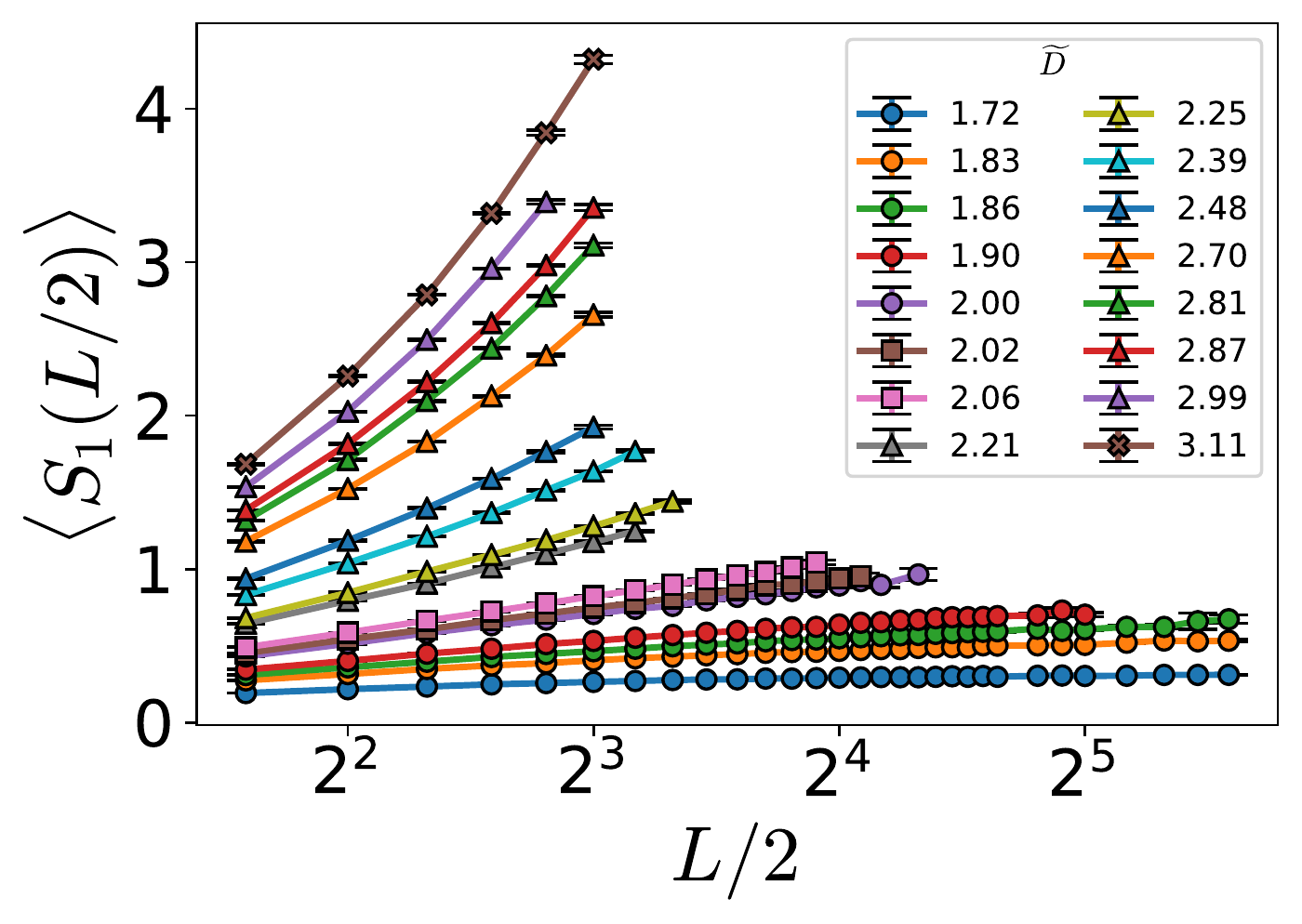} \label{fig:S1_vs_L}}
    \subfloat{\includegraphics[width=0.5\columnwidth]{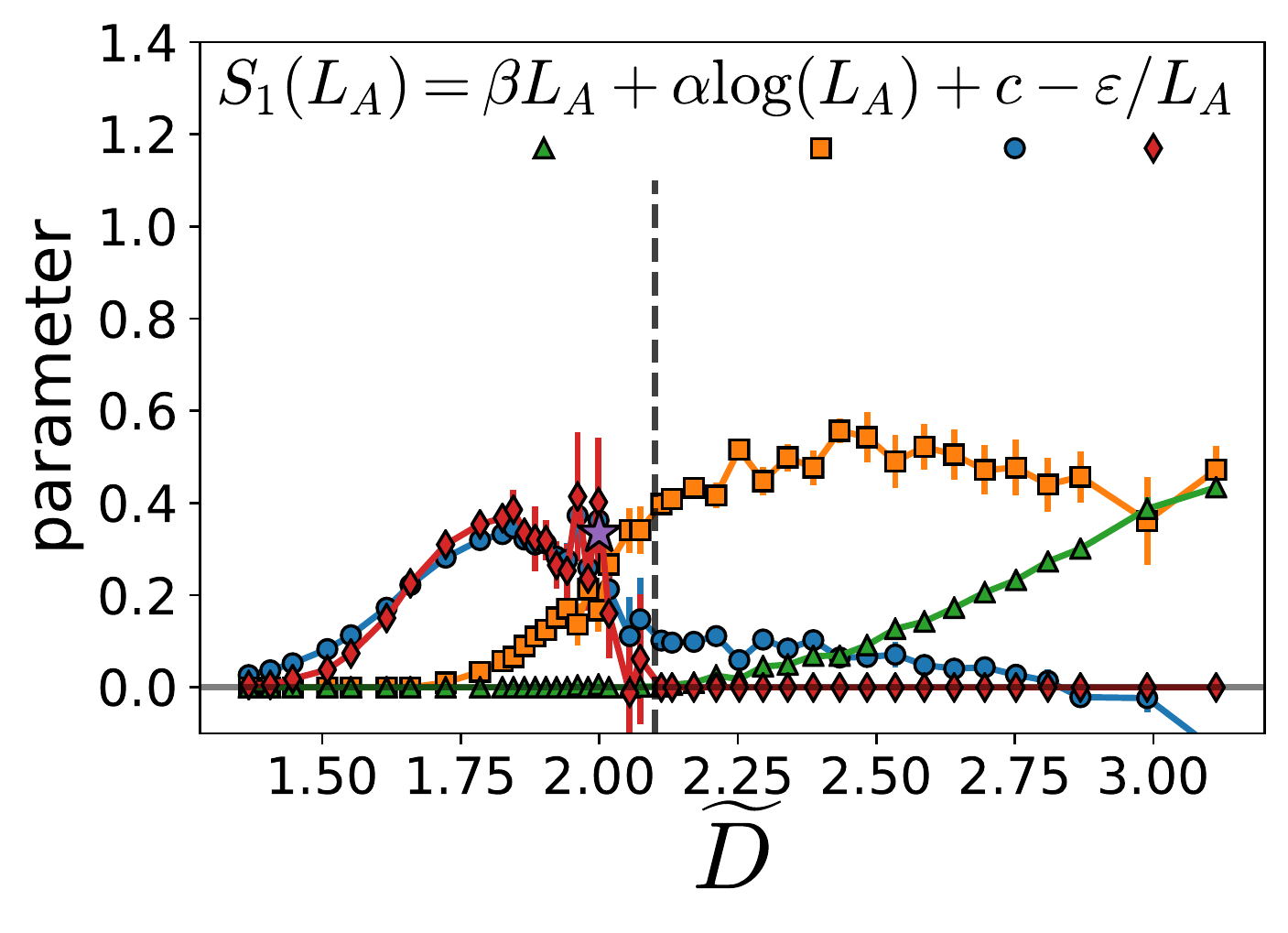}\label{fig:S1_vs_L_params}}
    \caption{\textit{Left:} Average von Neumann entanglement entropy $\langle S_1(L/2) \rangle$  for a non-uniform network for various values of $\dt$ on a semi-log plot. 
    \textit{Right:} parameters when fitting $\langle S_1(L/2)\rangle$ to the functional form $\langle S_1(L_A=L/2) \rangle  = \beta L_A + \alpha \log L_A + c - \varepsilon/L_A$ where $\alpha,\beta,c,\varepsilon$ are determined by a non-linear least-squares fit excluding the last two data points. We use $\alpha,\beta \geq 0$ and in order to ensure convergence of the fitting routine $\varepsilon=0$ for $\dt > 2.1$, shown as a dashed line.}
    \label{fig:S1_fits}
\end{figure}

We now consider the mean $\langle S_n(L/2) \rangle$ as a function of $L$ directly for the non-uniform tensor networks (see fig.~\ref{fig:S1_vs_L_params} left and supplementary fig.~\ref{fig:S1_examples}). We visually see that at $\dt\gtrsim 2.25$ the entropy $S_n(L/2)$ grows faster then logarithmically and at $\dt\approx 2.0$ scales logarithmically over the whole range of $L$. At small $\dt < 1.9$ the entanglement entropy appears concave on the semi-log plot as $L$ increases.

To better quantify this, we model the relationship $\langle S_1(L/2)\rangle$ by fitting the entropy to a functional $\langle S_1(L_A=L/2)\rangle = \beta L_A + \alpha \log(L_A) + c - \varepsilon/L_A$ where both $\alpha$ and $\beta$ are restricted to be non-negative so as to make sense physically (see fig.~\ref{fig:S1_vs_L_params} right)).
To ensure good fitting convergence, we also force $\varepsilon=0$ after $\dt>2.1$, as the small $L$ corrections are less prominent than  $\dt < 2.1$.
At extremely small values of $\dt\approx 1.55$, we  are able to see entropy that is independent of $L$ at large $L$ saturating at the fitted constant $c$.  For $1.55<\dt<1.7$, we never reach system sizes which saturate the entanglement but our fit shows that $\alpha=0$ and hence there is no need for a logarithmic term to explain the data indicating that this region is area law. 

Between $1.7 < \dt < 2.05$ there is a consistent logarithmic term to the fitted data with no linear term.  Interestingly at $\dt=2.06$ the linear and $1/L$ fit are nearly zero and we get essentially a perfect logarithmic fit as in the uniform case. The measured slope of the logarithm term (fit only to $S_1(L/2)=\alpha \log (L)$) is $0.291(2)$ at $\dt \approx 2.0$ and $0.33$ at $\dt=2.05$ (which matches to the corresponding $D=2$ uniform slope).   Having an extended region of logarithmic scaling could be the result of either a separate logarithmic scaling phase or a critical fan around a logarithmic critical point.  

In the region $\dt \gtrsim 2.1$, we observe that linear coefficient $\beta$ becomes non-zero with a roughly constant subleading logarithmic term.  
A volume law accompanied with a stable logarithmic term  agrees with the original observations of refs~\cite{Li2019,PhysRevB.99.224307}.  Other recent results~\cite{Li2021} find that the subleading term for Clifford circuits within the volume law phase goes as $L^{1/3}$.  While for any given $\dt$ it is hard to distinguish a subleading logarithm vs power-law, we do find the best-parameter fit for the logarithmic subleading correction is smoother as a function of $\dt$ then those found by fitting $\dt^{1/3}$.  

To further identify where the linear term vanishes, we measure the derivative of the entropy $d\langle S_1\rangle /dL$ (see  fig.~\ref{fig:derivative}) and determine where it goes to zero in the infinite system limit (i.e. $L\rightarrow \infty$).  A zero derivative in this limit captures the transition point from area or log law (zero slope) to volume law (non-zero slope). Statistical noise, partially due to increasing variance per site  (see  fig.~\ref{fig:Sn_stats}), makes it hard to precisely measure the derivative. That said, despite this not being particularly robust, the derivative extrapolation gives a transition at $\dt = 2.03$.  
  
\subsubsection{Entanglement Entropy Distributions}
\begin{figure}
    \centering
    \includegraphics[width=\columnwidth]{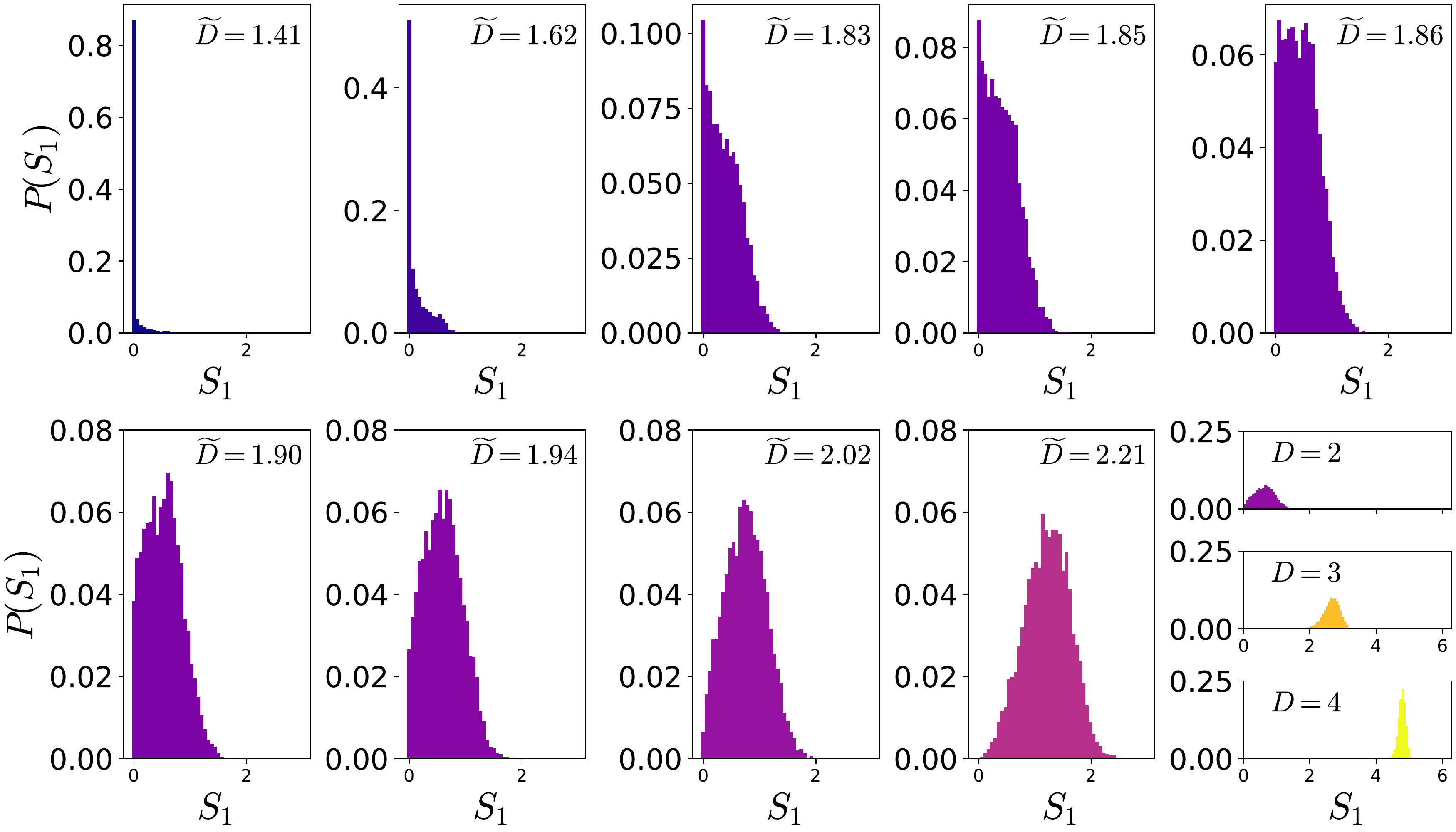}
    \caption{Histograms of the von Neumann entanglement entropy $S_1(L/2)$ for various $\widetilde{D}$ with a system of width $L=18$ and bins of size 0.06. \textit{Lower right:} uniform network distributions with $L=10$ and bins of size 0.06}
    \label{fig:S1_histograms_dist}
\end{figure}

\begin{figure}
    \centering
    \subfloat{\includegraphics[width=0.5\columnwidth]{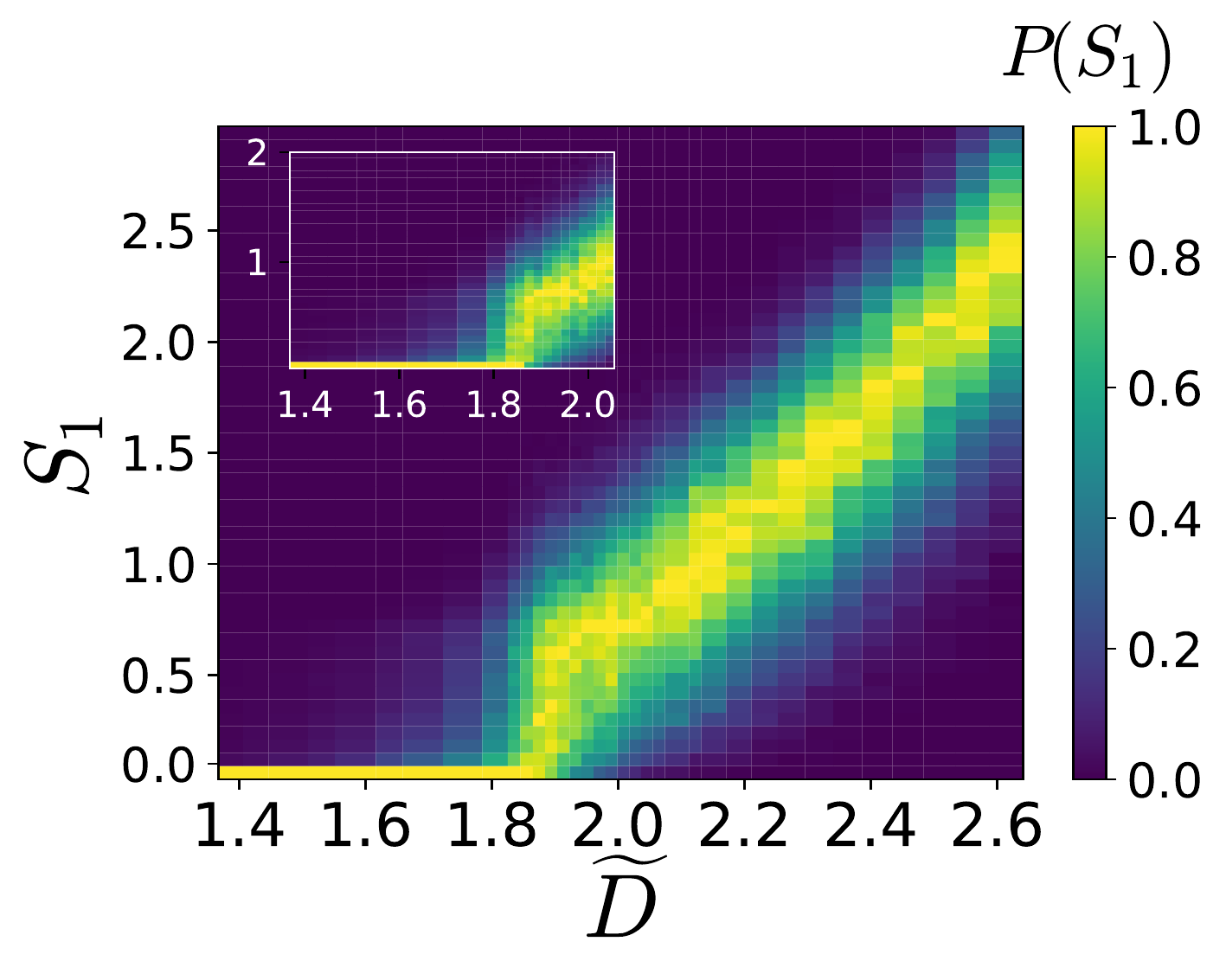}\label{fig:hist_vs_D}}
    \subfloat{\includegraphics[width=0.55\columnwidth]{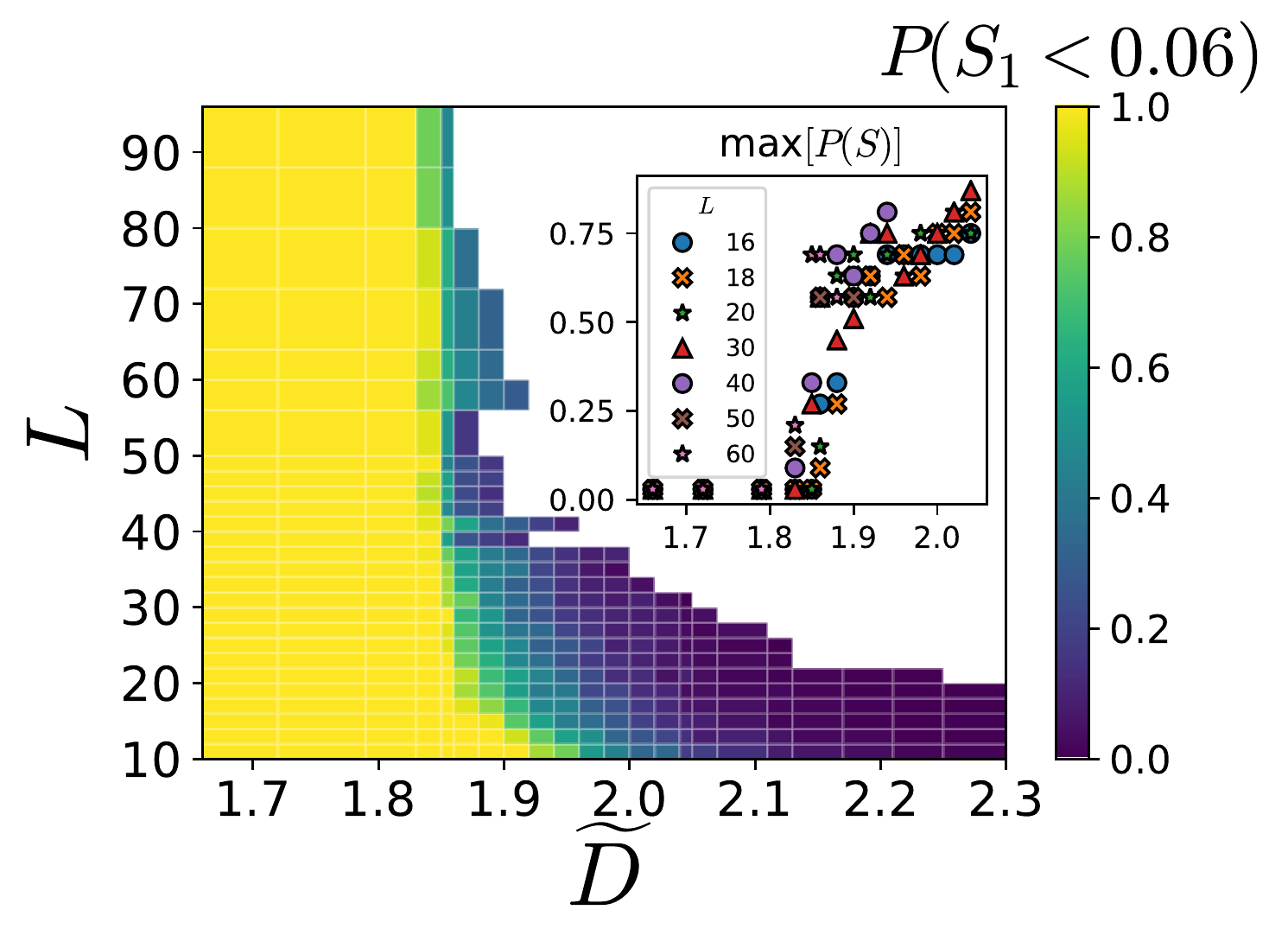}\label{fig:min_hist_L_vs_D}}
    \caption{\textit{Left:} normalized histogram of the von Neumann entanglement entropy $S_1(L/2)$ as a function of $\dt$ for a system of width $L=16$ (\textit{inset:} $L=28$) with bins of width 0.06. The histogram suggests a transition from a nearly pure area law phase to an entangled area law phase.
    \textit{Right:} system width $L$ dependence on $P(S_1<0.06)$ or nearly-zero entropy.  \textit{Inset:} mode of the binned entanglement entropy distribution as a function of $\dt$ for various $L$. }
    \label{fig:S1_histograms}
\end{figure}

We then examine the distribution of entanglement entropy $S_1(L/2)$ (see fig.~\ref{fig:S1_histograms_dist}) at several different $\dt$. There is a clear qualitative shift as one moves from low to high $\dt$.

At low $\dt$, we see an entropy distribution that is peaked at zero. As we go to larger $\dt$ we see a sharp change in the mode of the distribution from essentially zero (i.e. $S_1<0.06$) to non-zero. This suggests a distinction between a `nearly pure' tensor network and an `entangled' area law phase which can be distinguished by the plurality of disorder realizations having essentially no entanglement.
For $L=16$, this change happens at approximately $\dt=1.95(1)$  with the peak location growing with $\dt$ after this point (see fig.~\ref{fig:S1_histograms}).  We further study the $L$ dependence finding the $\dt$ at which the mode jumps recedes to smaller values saturating to around $\dt\sim 1.87$. 

While the $n=1$ data is presented here, the same qualitative features are found for the $n=2,\infty$ R\'enyi entropy shown in the supplementary. 

\subsubsection{Data Collapse}
\label{subsec:mutual_info}
To measure the critical exponents of the transition, we will employ two techniques: data collapse of the average entropy and the tripartite quantum mutual information (QMI).

First we directly compute the data collapse of $\langle S_1 \rangle$ (see fig.~\ref{fig:data_collapse}, and further details in the supplementary). Using this scaling ansatz, we find a critical point of $\dt_c=2.02(7)$ and critical exponent of $\nu=1.5(1)$. Extending this analysis to $n>1$, we find most values are around $\dt_c=2.04(2)$ and $\nu=1.58(7)$. Note that $\nu>1$ which satisfies the claim that the bond dimension disorder is irrelevant in our 2D tensor network. 

Another way to probe the transition is through quantum mutual information (QMI).  This has been suggested by refs.~\cite{Li2019,Gullans2019a,PhysRevB.101.060301} to be less sensitive to finite size effects. 
We take the approach of refs.~\cite{Gullans2019a,PhysRevB.101.060301} and focus on the tripartite mutual information $I_3(A:B:C)$ as it allows for studying smaller system sizes then looking at antipodal mutual information.  

To calculate $I_3$, we compute $I_3(A:B:C)=I^{(n)}(A:B)+I^{(n)}(B:C)-I^{(n)}(A:BC)$ where $I^{(n)}(A:B)=S_n({A})+S_n({B})-S_n({A,B})$ is the mutual information for groups of sites $A$ and $B$. We use regions $A,B,C$ as contiguous regions of size $L/4$ in the middle of the MPS, shown in the upper left of fig.~\ref{fig:data_collapse} right, so that the value of $I_3$ is the same under any permutation of $A,B,C$. Note that we calculate $I_3$ per layer, averaging over layers rather than using $\langle S_n \rangle$.

We compute the tripartite mutual information and use finite-size scaling \cite{pyfssa,HoudayerFiniteScaling} to obtain data collapse for $L=8,12,16,20,24$ shown in fig.~\ref{fig:data_collapse} right. 
We perform scaling analysis by slowly reducing the fit $\dt$ range and computing the mean from the produced set of potential $\dt_c,\nu$.
The collapse suggests a critical point at $\dt_c=2.03(3)$, with a critical exponent of $\nu=3.1(7)$.
Note that both $\nu=3.1(7)$ and $\nu=1.5(1)$ are distinct from the percolation exponent of $\nu=4/3$ \cite{stauffer1992introduction}. 
 
There are two potentially unphysical results when extending the tripartite QMI to $n>1$. The first is that the value of $\dt_c^{(n)}$ has a strong R\'enyi index dependence. Because $S_\infty$ bounds all $S_n$ for $n>1$ both above and below, all $n>1$ should simultaneously have the same critical point \cite{Li2019,PhysRevB.101.060301}. In addition, it is expected that $\nu$ is the same for all $n$, but we have a weaker but noticeable dependence on $n$.

\subsubsection{Decay of Pairwise Quantum Mutual Information}
Another quantity of interest is the QMI decay $\langle I^{(n)}(x)\rangle$ as a function of distance $x$ between two sites. This quantity decays
for Haar-random circuits
as $I^{(n)} \propto 1/x^{2\Delta_c}$ at the transition with $\Delta_c=2$ as expected from percolation with periodic boundary conditions \cite{PhysRevX.9.031009}. 
We show $\langle I^{(1)}(x)\rangle$ at $x=L/2$ as a function of $L$ in fig.~\ref{fig:qmi_decay}. In a fairly large range of $1.7 \leq \dt \leq 2.3$ the QMI appears to decay algebraically as $1/x^3$.  Both the cubic as opposed to quartic rate of the algebraic decay at the transition and the fact that we don't see clear exponential decay outside (albeit still near) the transition is interesting.  

\begin{figure}
    \centering
     \subfloat{\includegraphics[width=0.5\columnwidth]{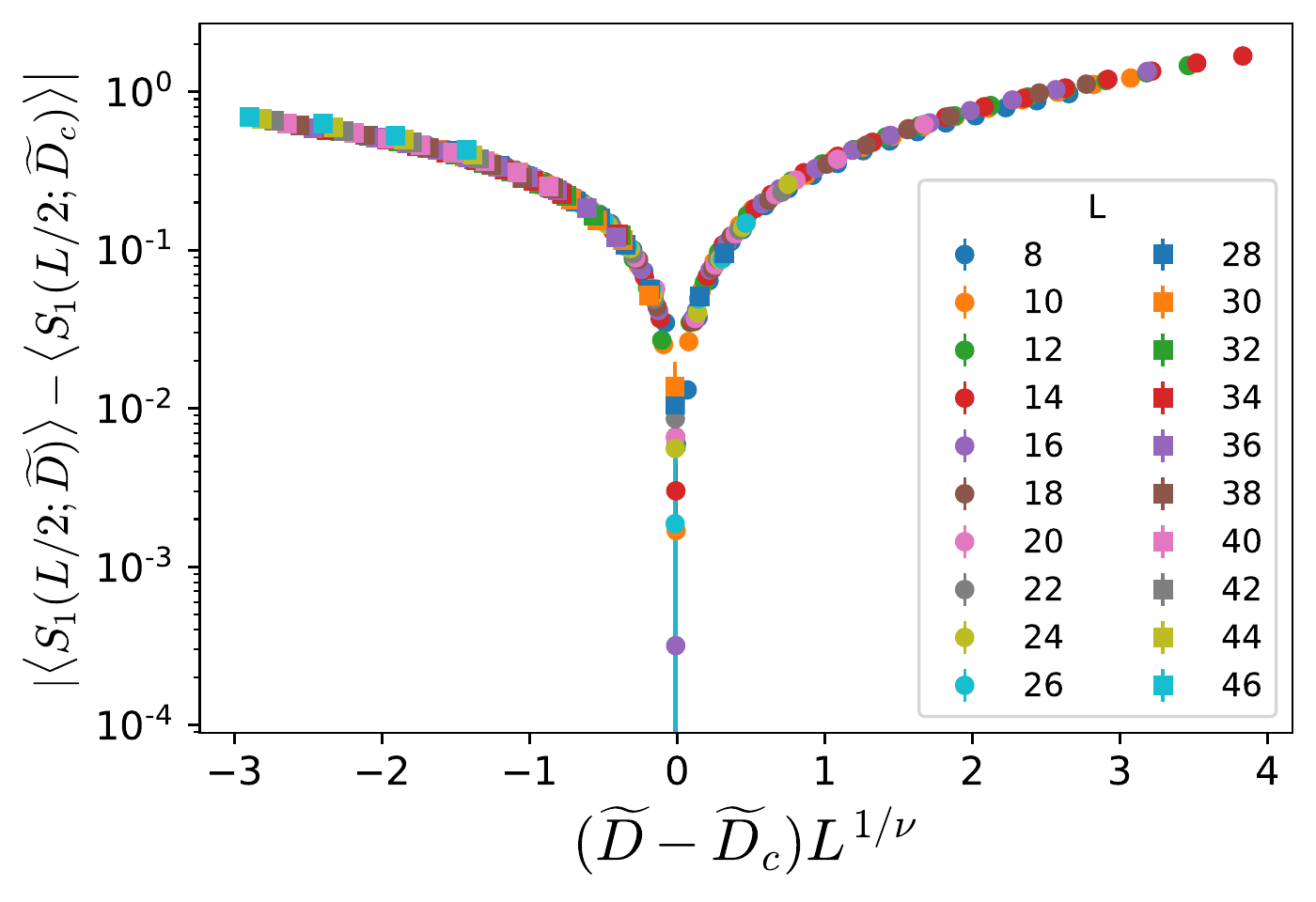}}
    \subfloat{\includegraphics[width=0.5\columnwidth]{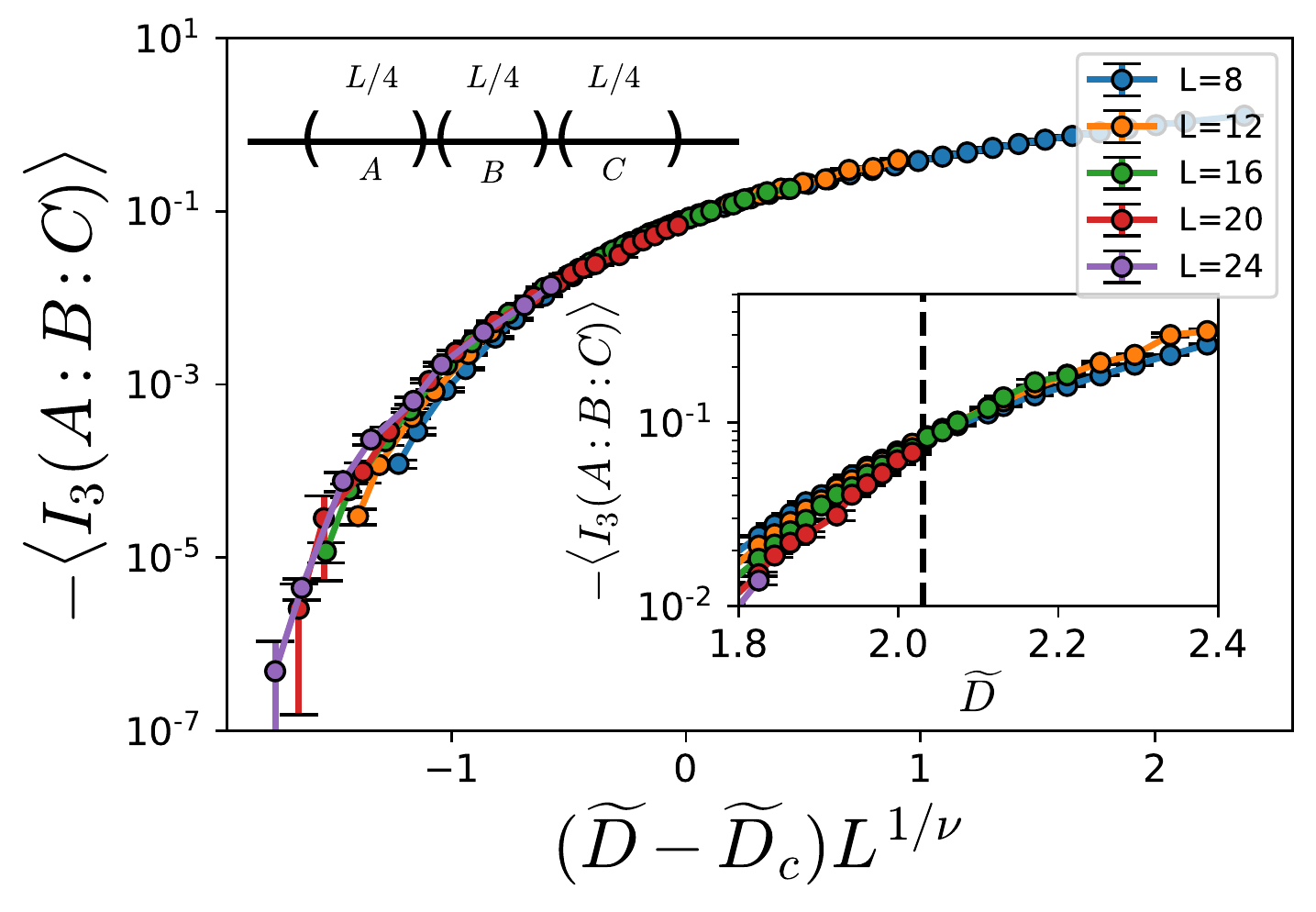}}
    \caption{ \textit{Left:} Scaling collapse of the von Neumann entanglement entropy $\langle S_1 (L/2)\rangle$ using a functional form $\Delta S_1 = f[(\dt -\dt_c)L^{1/\nu}]$ for $\Delta S_1 = |\langle S_n(L/2;\dt) \rangle - \langle S_n(L/2;\dt_c) \rangle|$, some universal function $f$, and $\dt_c=2.02(7)$, $\nu=1.5(1)$. 
    \textit{Right:} Scaling collapse of $I_3(A:B:C)$, the tripartite mutual information in the center of the MPS (depicted in the upper left) for $L=8,12,16,20,24$. We find critical exponents $\widetilde{D}_c= 2.03(3)$ and $\nu=3.1(7)$ using a scaling ansatz of $I_3 = f\left[(\dt -\dt_c)L^{1/\nu} \right]$ for some universal function $f$. 
    \label{fig:data_collapse}}
    
\end{figure}

\begin{figure}
    \centering
    \includegraphics[width=\columnwidth]{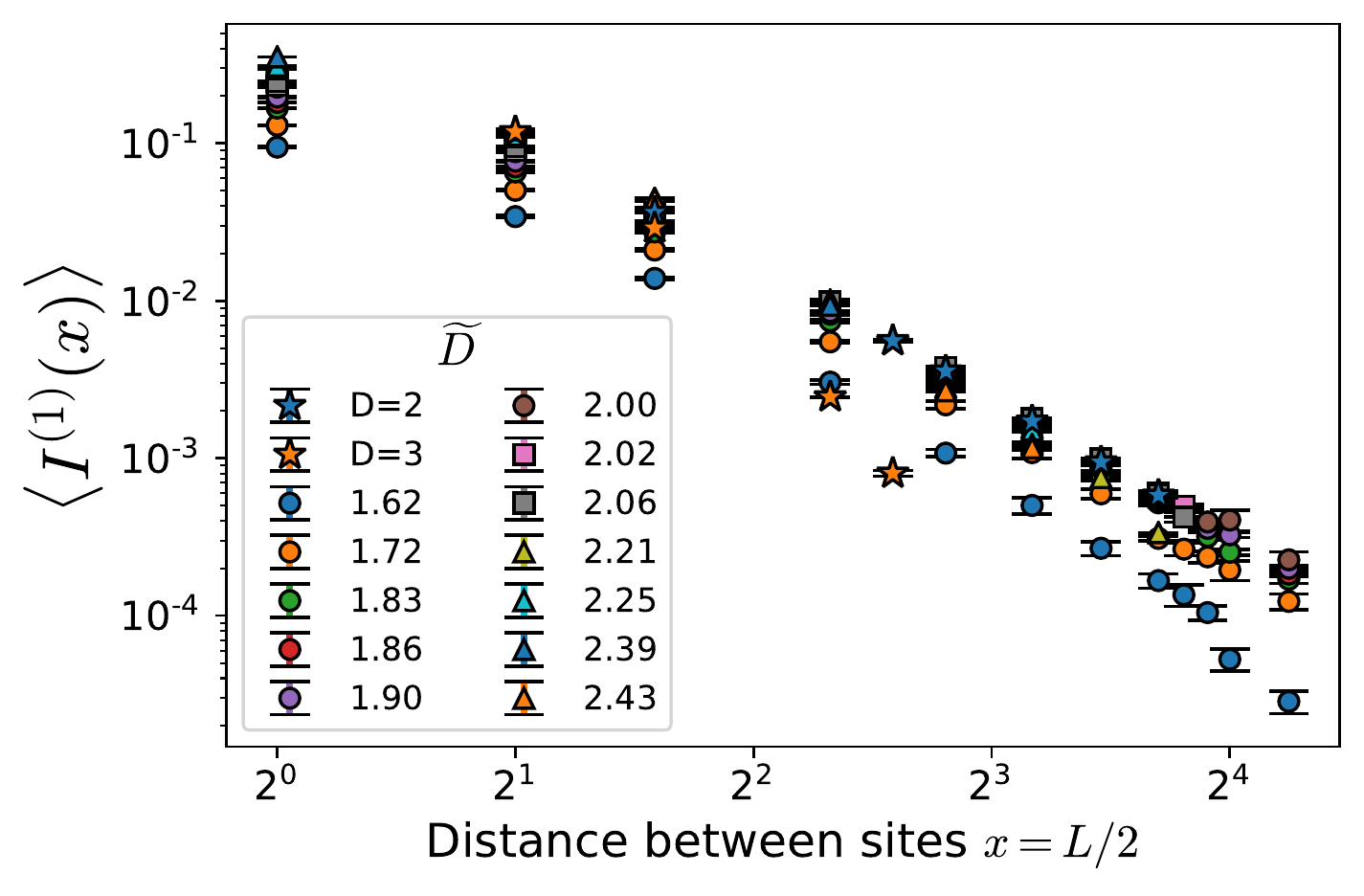}
    \caption{ Mutual information $\langle I^{(1)}(x)\rangle$ between two sites located a distance $x=L/2$ away for various values of $\dt$. $D=3$ and $\dt=1.67$ appears to show exponential decay, while most other data shows algebraic decay.   }
    \label{fig:qmi_decay}
\end{figure}

\section{Conclusion}

We study a tensor network with both uniform and log-normal distributed bond dimensions. The uniform network has logarithmic behavior at $D=2$ , consistent with a critical point predicted in \cite{Vasseur2018}, and follows with a volume law regime at large uniform $D$ as predicted. We also find a logarithmic slope coefficient of about $1/3$  at $D=2$ matching the analysis of ref.~\cite{Jian2019}. 

Utilizing a disordered network, we study in more detail the behavior around $D=2$. The disordered network recovers the same behavior in the logarithmic term as the uniform network with an appropriate shift ($D=2 \leftrightarrow \dt\approx 2.06$), and in the volume law phase has a persistent sub-leading correction, most likely logarithmic. Further understanding of this subleading behavior is warranted for future study.

We verify the location of the transition and the critical exponent $\nu$. Using the noisy $d\langle S_1(L/2)\rangle/dL$ shows an extrapolated critical point at $\dt=2.03$, and using data collapse methods show $\dt^S_c=2.02(7)$ and $\dt^{QMI}_c=2.03(3)$ which all agree within error bars. The entropy and tripartite QMI methods disagree on $\nu$ however, where the methods measure a value of $\nu=1.5(1)$ and $\nu=3.1(7)$ respectively. We also see using the tripartite QMI R\'enyi index dependence  on the critical exponent, in both the $\dt_c$ and more weakly in $\nu$, where for the entanglement entropy directly it's roughly constant at an average of $\nu=1.58(7)$ for $n>1$. 

In all of these cases, however, the slope of the logarithmic term at these critical points are not the universal $1/3$ predicted in ref.~\cite{Jian2019}, but does follow the CFT inspired ansatz of ref.~\cite{PhysRevB.101.060301}.
We find that our critical exponents differ significantly from those found in other works~\cite{PhysRevB.101.060301,Li2019,Bao2019} looking at quantum circuits and random tree tensor networks suggesting that the rectangular geometry random tensor network model transition is in a different universality class.

Additionally the network has unanticipated behavior both away from and at the transition. Using entanglement entropy distributions we find a change in the low entropy behavior of the network as a signal of `nearly pure' vs `entangled' area law states, reminiscent of the area law phase of random quantum circuits \cite{Gullans2019a} around $\dt \sim 1.87$. A non-zero leading logarithmic term between $\dt\approx 1.7-2.05$ and algebraic decay of the pairwise mutual information around $\dt \approx 1.9-2.1$ additionally give some hints of a possible log-law phase around the critical point.  It is also interesting that the QMI scaling collapse for $n\ge 1$ identifies a transition at $\dt=1.81(1)$.   However it is entirely possible these are coming from a critical fan or additional finite size effects and not a separate phase. The pairwise mutual information decay also goes as  $I_1\propto 1/x^3$ unlike the $1/x^4$ of the quantum circuits.  

Interesting future work on this model may include characterizing other critical exponents which have clear probes but are difficult to simulate for tensor networks with large entanglement \cite{Gullans2020,PRXQuantum.2.030313}.  
Other directions may come from the broader tensor network community, where improved tensor network simulation techniques for contracting PEPS \cite{peps2020pang} and improved parallelization \cite{dmrg2020RL} may benefit our understanding.
The random tensor network is a paradigmatic model for entanglement transitions and further study of this system will help elucidate
properties of the broader class of entanglement transitions. 

\section{Acknowledgements}
We thank R. Vasseur, A. W. W. Ludwig, and T. Hsieh for useful discussions. 
We acknowledge support from the Department of Energy grant DOE DESC0020165.
This research is part of the Blue Waters sustained-petascale computing project, which is supported by the National Science Foundation (awards OCI-0725070 and ACI-1238993) the State of Illinois, and as of December, 2019, the National Geospatial-Intelligence Agency. Blue Waters is a joint effort of the University of Illinois at Urbana-Champaign and its National Center for Supercomputing Applications. This work made use of the Illinois Campus Cluster, a computing resource that is operated by the Illinois Campus Cluster Program (ICCP) in conjunction with the National Center for Supercomputing Applications (NCSA) and which is supported by funds from the University of Illinois at Urbana-Champaign.
\bibliography{randomMPO}

\newpage
\clearpage
\appendix
\section{Supplementary Material}
\renewcommand{\thefigure}{S\arabic{figure}}
\renewcommand{\theequation}{S\arabic{equation}}

\setcounter{figure}{0}
\setcounter{equation}{0}

\begin{figure}[t]
    \centering
    \includegraphics[width=\columnwidth]{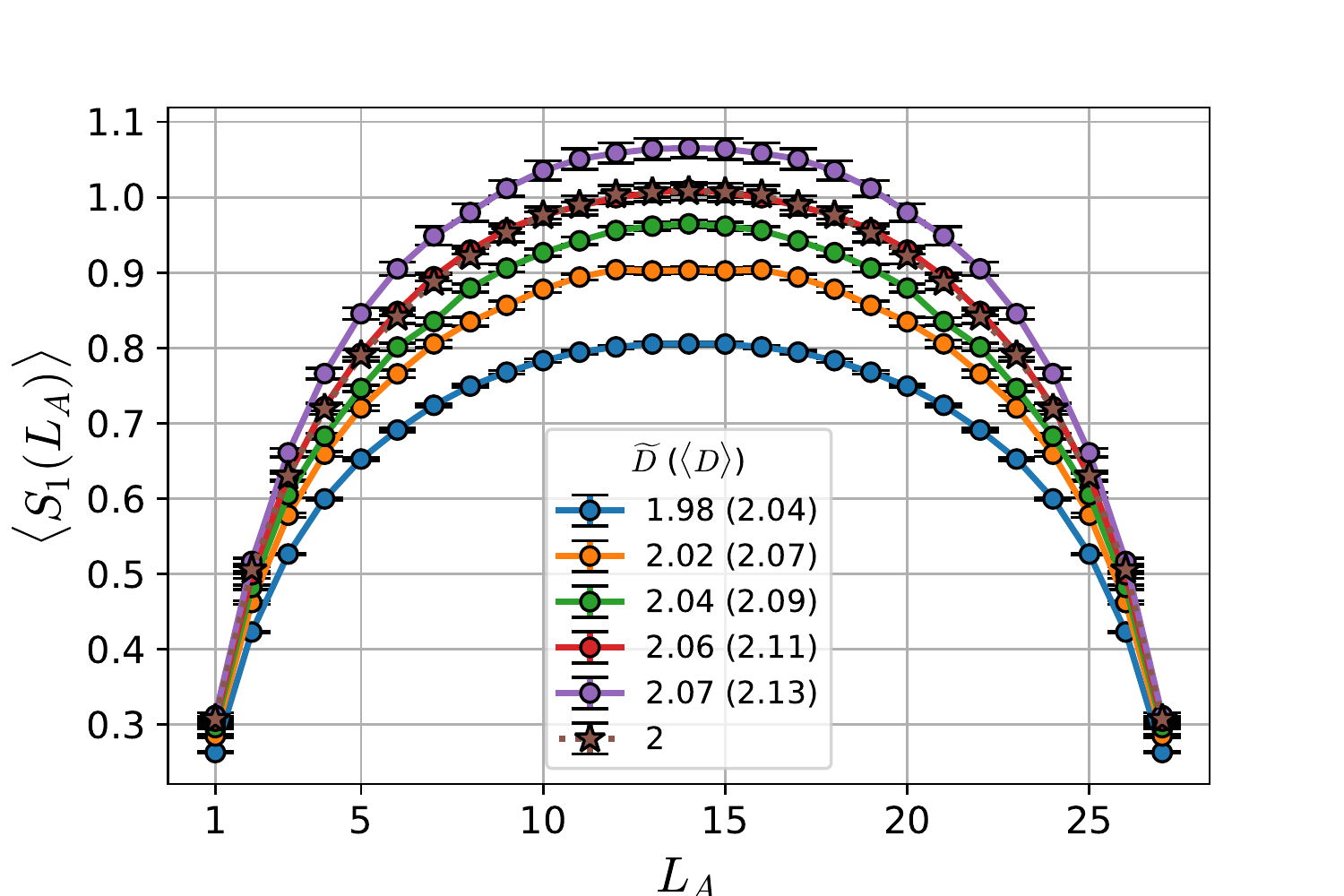}
    \caption{Comparison of the entanglement entropy $\langle S_1(L_A)\rangle$ of a width $L=28$ system at different cuts $L_A$ for various $\dt$. Note that $\dt = 2.05$ and $D=2$ have nearly the same curve. }
    \label{fig:dtilde_comparison}
\end{figure}

\begin{figure}[t]
    \centering
    \includegraphics[width=0.9\columnwidth]{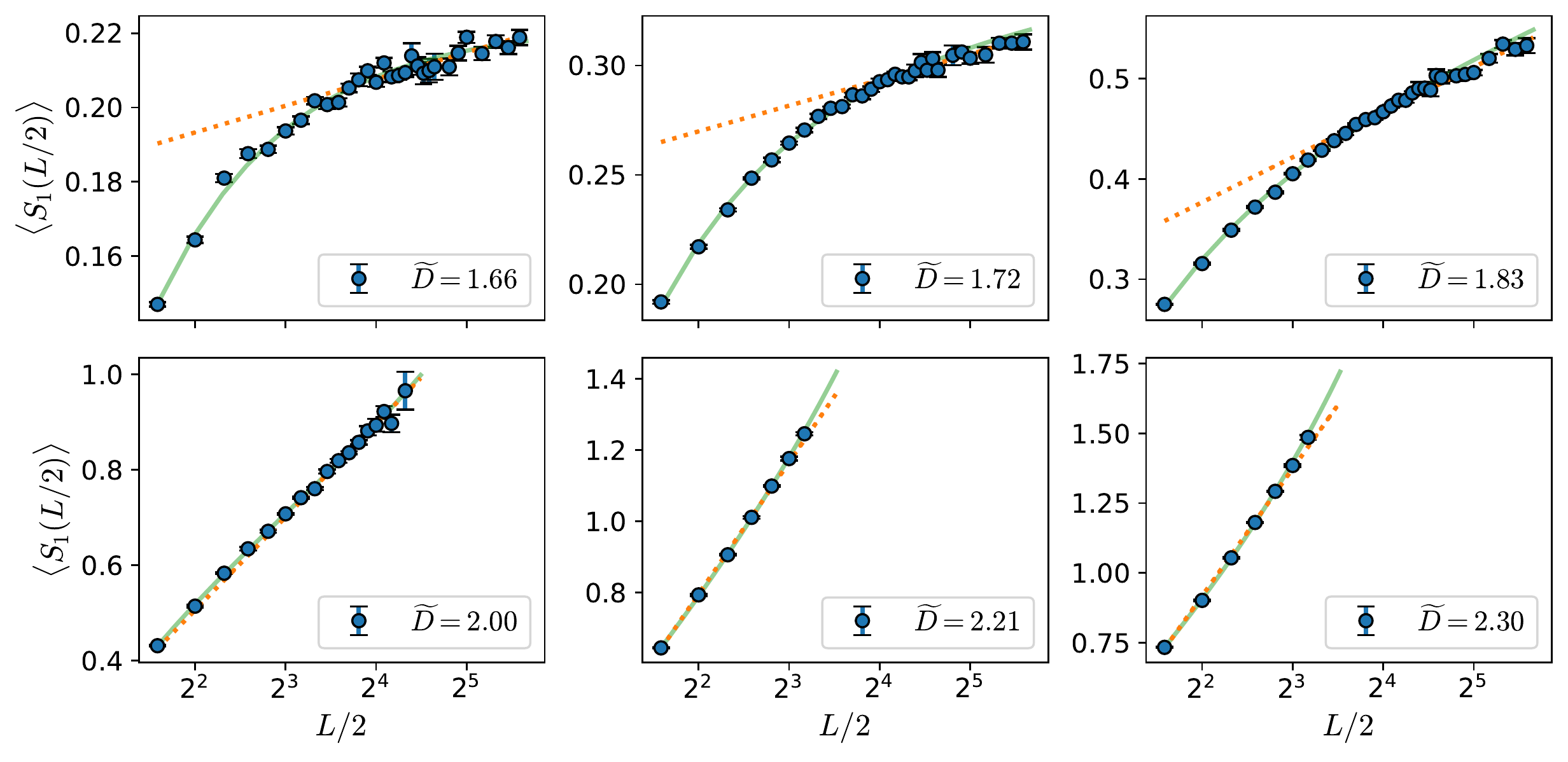}
    \caption{Comparison of the von Neumann entanglement entropy $\langle S_1(L/2)\rangle$ at various system widths for various $\dt$. The solid line represents a functional fit as in fig.~\ref{fig:S1_fits} and the dotted line is a logarithmic fit; both fits utilize the largest 20 system sizes accessible.  }
    \label{fig:S1_examples}
\end{figure}

\subsection{Higher R\'enyi Entropy Statistics}

\begin{figure}
    \centering
    \includegraphics[width=\columnwidth]{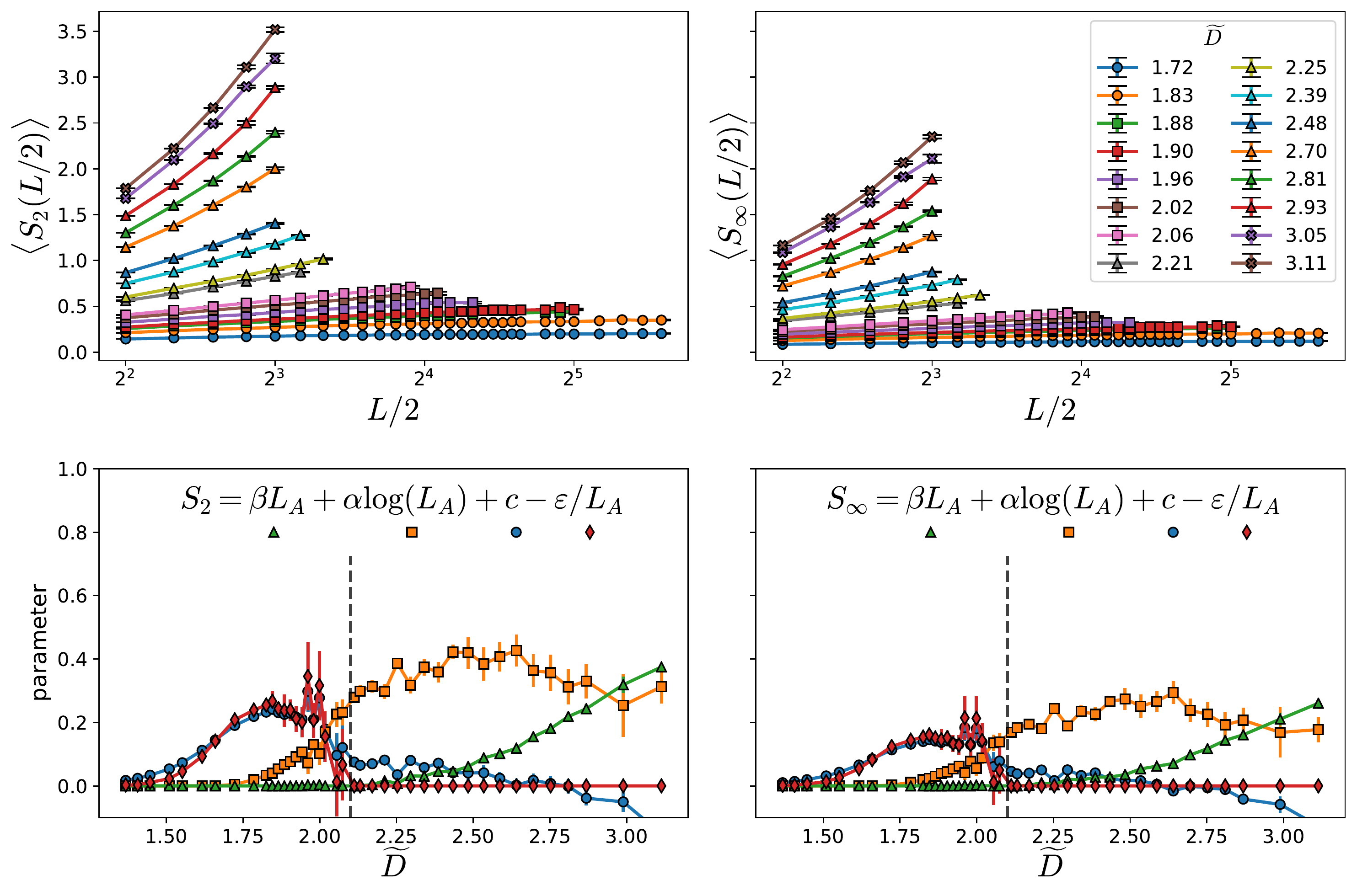}
    \caption{\textit{Top row:} Average entanglement entropy $\langle S_n (L/2)\rangle$ for different R\'enyi entropy indices $n=2$ (left), $n=\infty$ (right) and various values of $\dt$ on a semi-log plot. 
    \textit{Bottom Row:} parameters when fitting $\langle S_n(L/2)\rangle$ to the functional form $\langle S_n(L_A=L/2) \rangle  = \beta L_A + \alpha \log L_A + c - \varepsilon/L_A$ where $\alpha,\beta,c,\varepsilon$ are determined by a non-linear least-squares fit excluding the last two data points. We use $\alpha,\beta \geq 0$ and in order to ensure convergence of the fitting routine $\varepsilon=0$ for $\dt > 2.1$, shown as a dashed line.}
    \label{fig:S_L_fits}
\end{figure}

We show the same data as fig.~3  for $n=2,\infty$ in fig.~\ref{fig:S_L_fits}. As we also see in the distribution data, the R\'enyi entropy behavior is consistent but shifted across the different values of $n$.

\begin{figure}
    \centering
    \subfloat{\includegraphics[width=0.5\columnwidth]{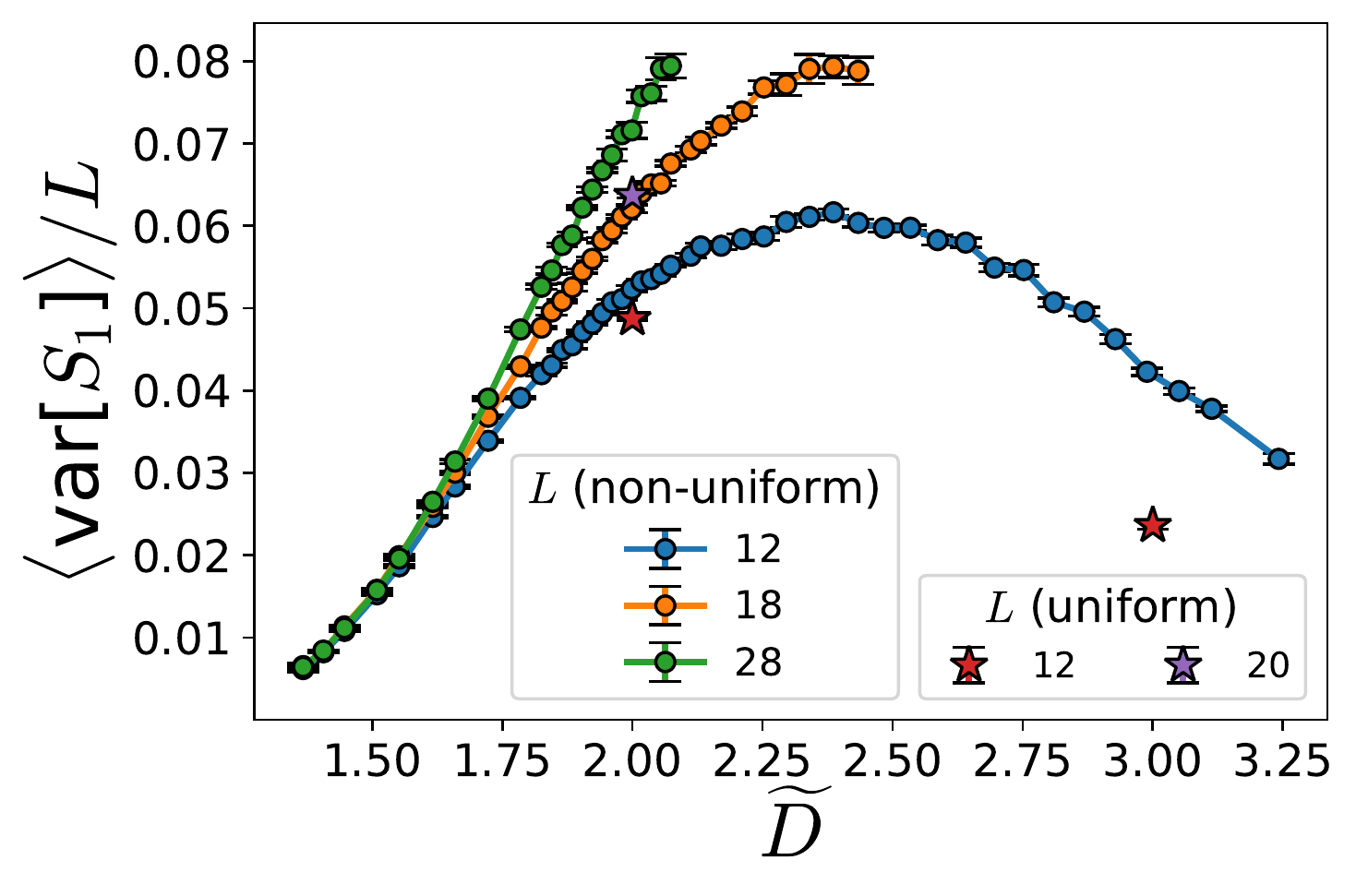}}
    \subfloat{\includegraphics[width=0.5\columnwidth]{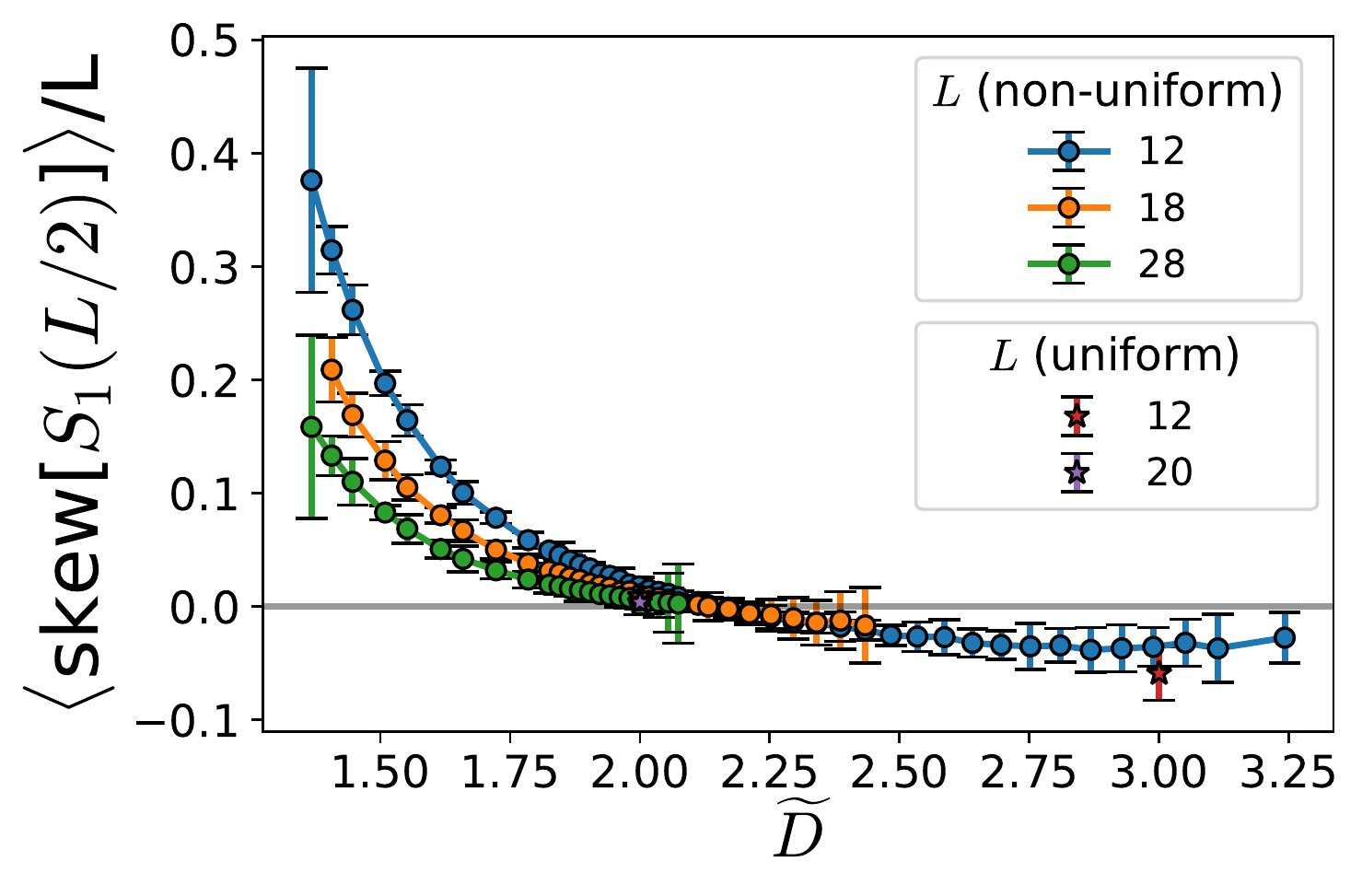}}
    
    \subfloat{\includegraphics[width=0.5\columnwidth]{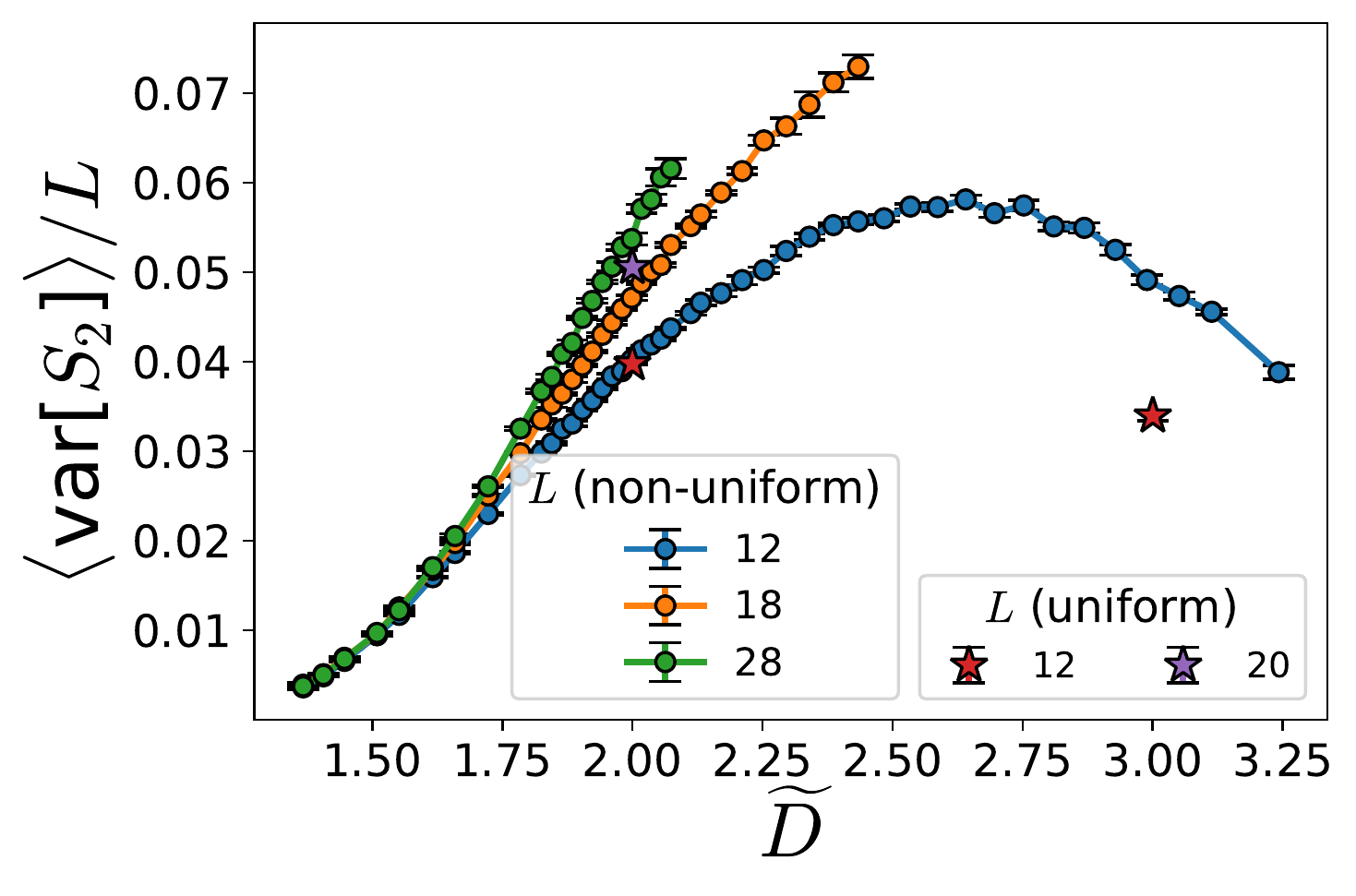}}
    \subfloat{\includegraphics[width=0.5\columnwidth]{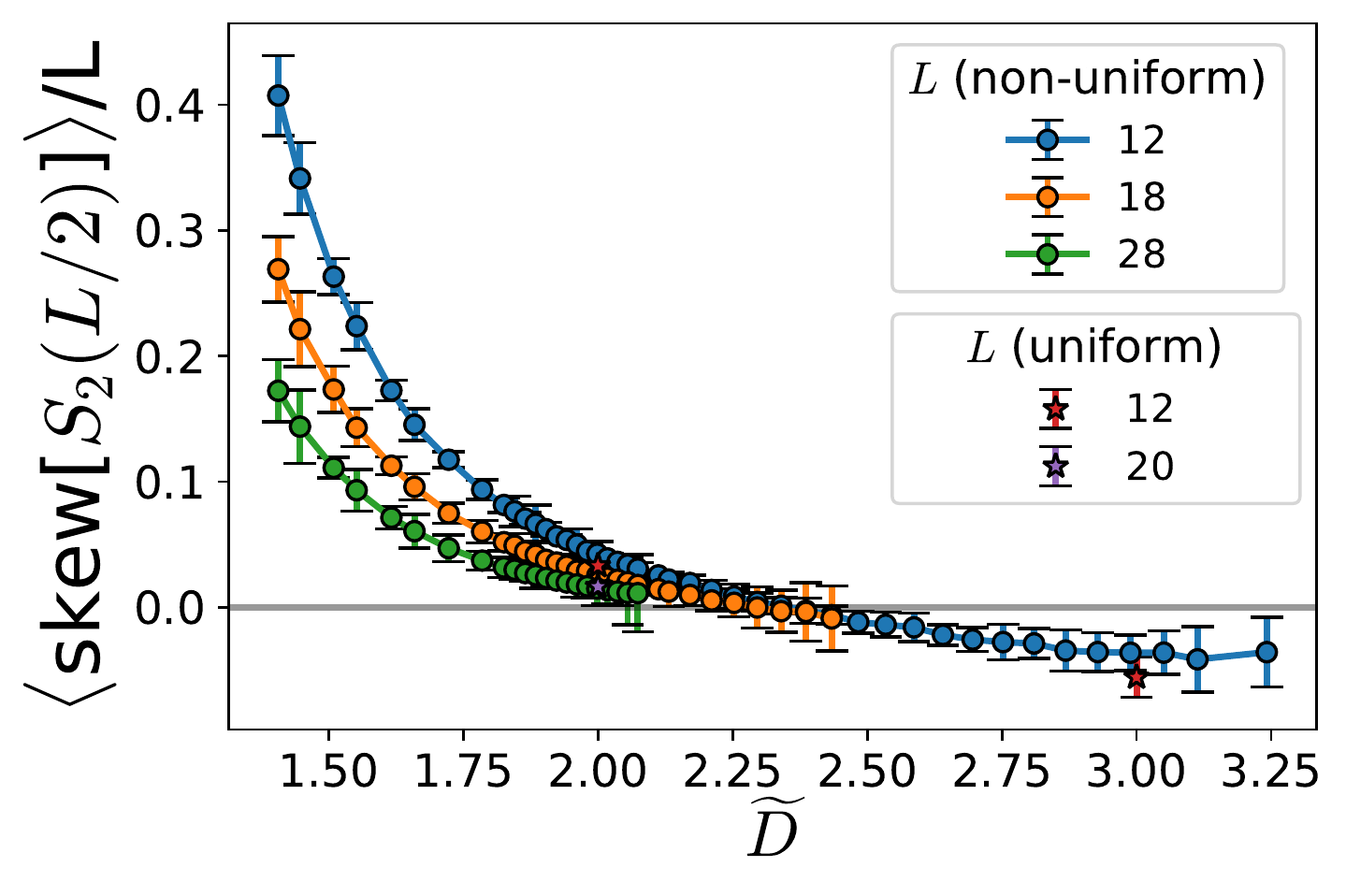}}
    
    \subfloat{\includegraphics[width=0.5\columnwidth]{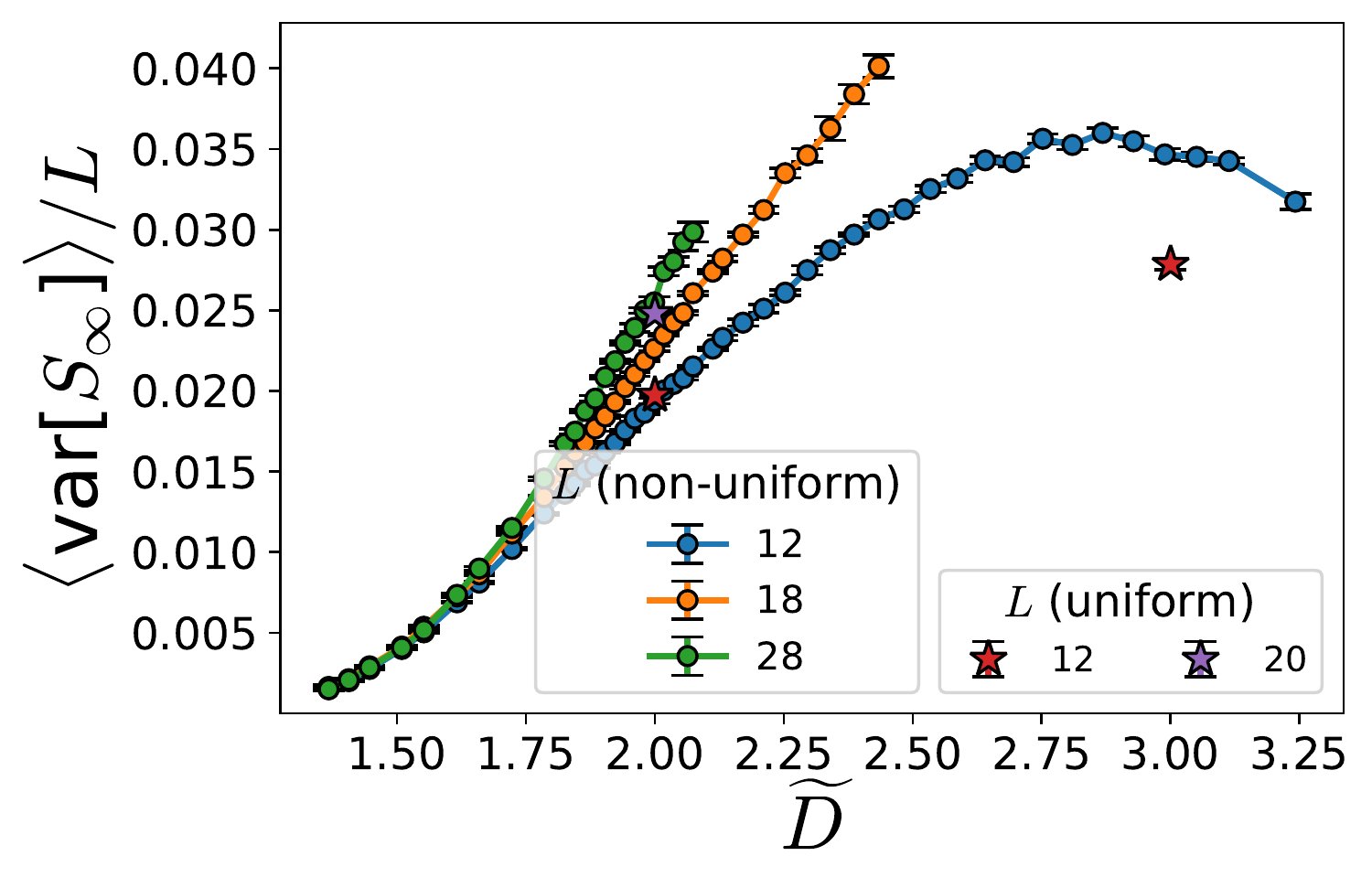}}
    \subfloat{\includegraphics[width=0.5\columnwidth]{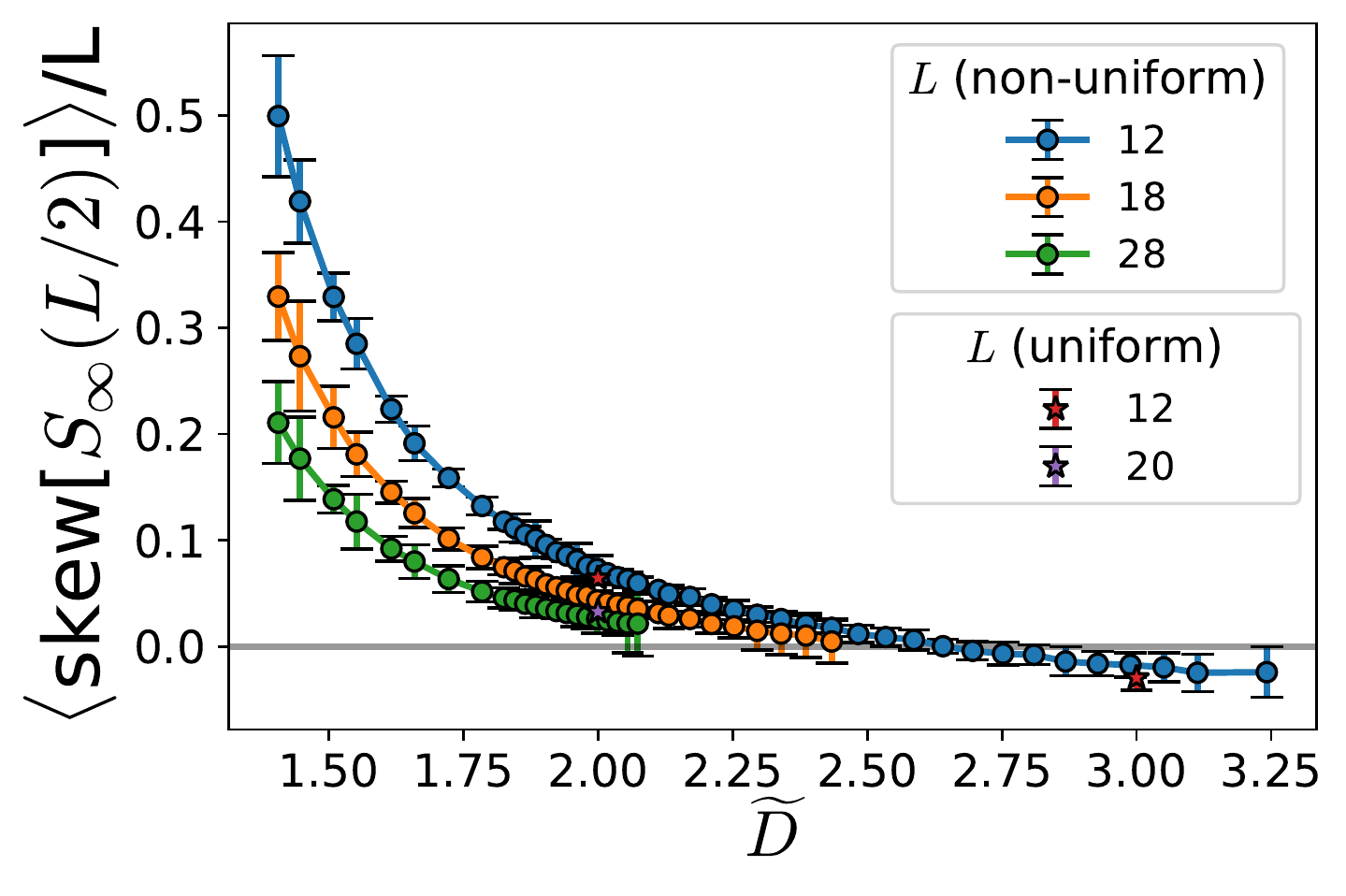}}
    
    \caption{ Value of the average skew (\textit{right}) and variance per site (\textit{left}) of the $n=1$ (top), $n=2$ (middle), and $n=\infty$ (bottom) R\'enyi entanglement entropy $\langle S_n(L/2)\rangle$ as a function of $\dt$ for systems of various width $L$, with uniform $D$ results shown with stars.}
    \label{fig:Sn_stats}
\end{figure}

Another signal for the transition may be found in the higher moments of entropy \cite{villalonga2020characterizing}. We show the variance and skew of the entropy for different values of $n$ in fig.~\ref{fig:Sn_stats}. Here we see that the skew goes from positive skew (indicating a strong number of low entropy values, consistent with the area law phase), to a negative skew at $\dt = 2.19(4)$ when fitting the average skew vs $\dt$ with a univariate spline and averaging over different system sizes $L$. In the limit of $\dt\to\infty$ we expect the distribution to return to a skew of 0. 
The variance appears to peak at intermediate $\dt$ potentially signalling the transition.  We find a significant shift in that peak as $L$ grows and given computational constraints are unable to locate the peak for the largest $L$.

Finally we look at the coefficient of a log fit at the critical point, for various R\'enyi entropy in fig.~\ref{fig:alpha_function}. Rather than track $\dt_c(n)$ we fix $\dt_c=2.030$ using a spline interpolation of the data.
Measuring the slope at the critical point gives an approximate functional form of $0.206(1+1/n)-0.102$; this functional form (with different constants) was found in ref.~\cite{PhysRevB.101.060301} for quantum circuits. We also find that the slope for the uniform $D=2$ matches ref.~\cite{Jian2019}

\begin{figure}
    \centering
    \subfloat{\includegraphics[width=\columnwidth]{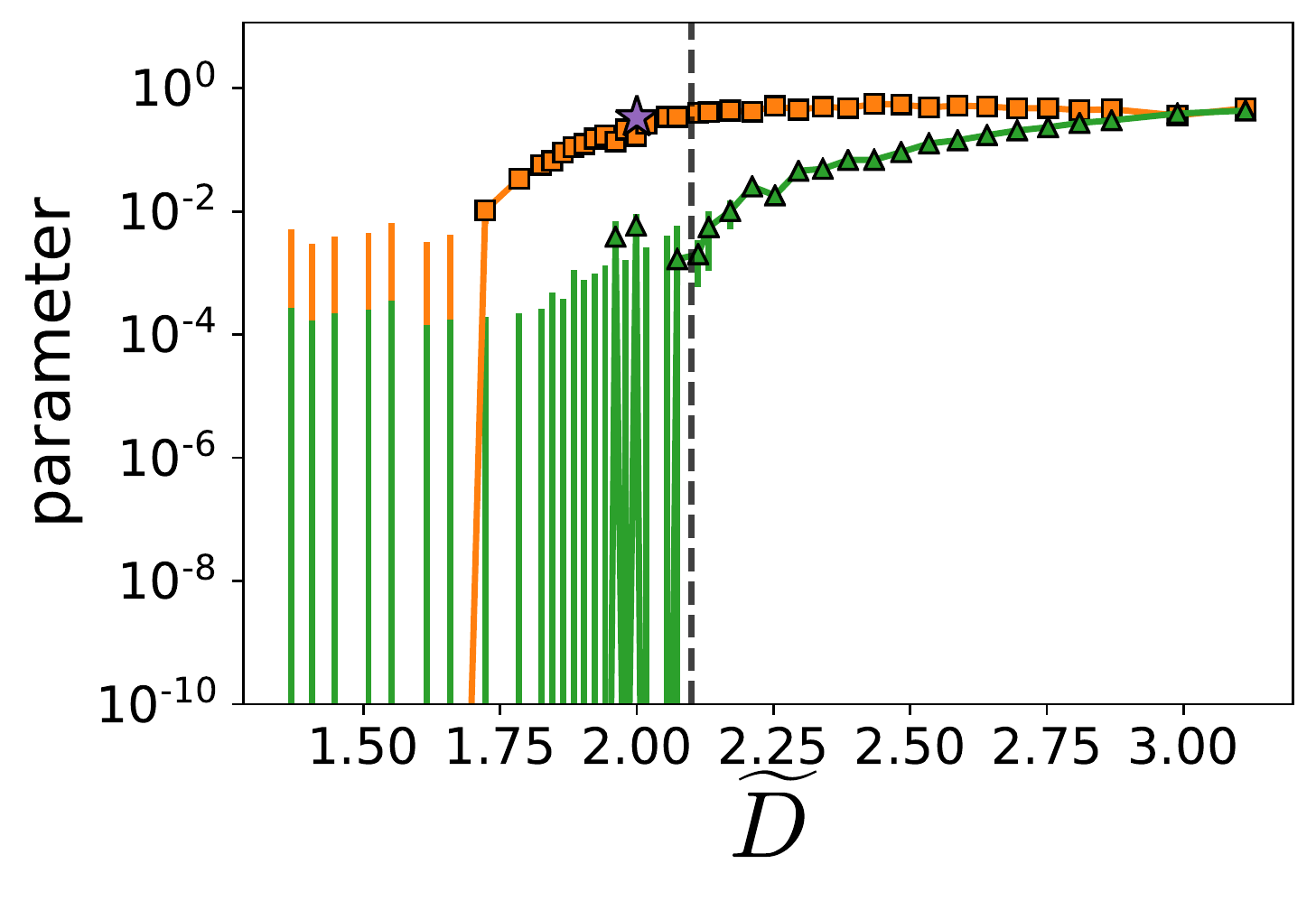}}

    \caption{Zoomed in parameters from fig.~3 on a semi-log scale. $\langle S_1(L/2)\rangle$ is fit to the functional form $\langle S_1(L_A=L/2) \rangle  = \beta L_A + \alpha \log L_A + c - \varepsilon/L_A$ where $\alpha,\beta,c,\varepsilon$ are determined by a non-linear least-squares fit excluding the last two data points. Shown are the linear (triangles) and log (squares) coefficients; a purple star marks the log slope for the uniform $D=2$.}
    \label{fig:S1_vs_L_params_zoom}
\end{figure}

\begin{figure}
    \centering
    \includegraphics[width=\columnwidth]{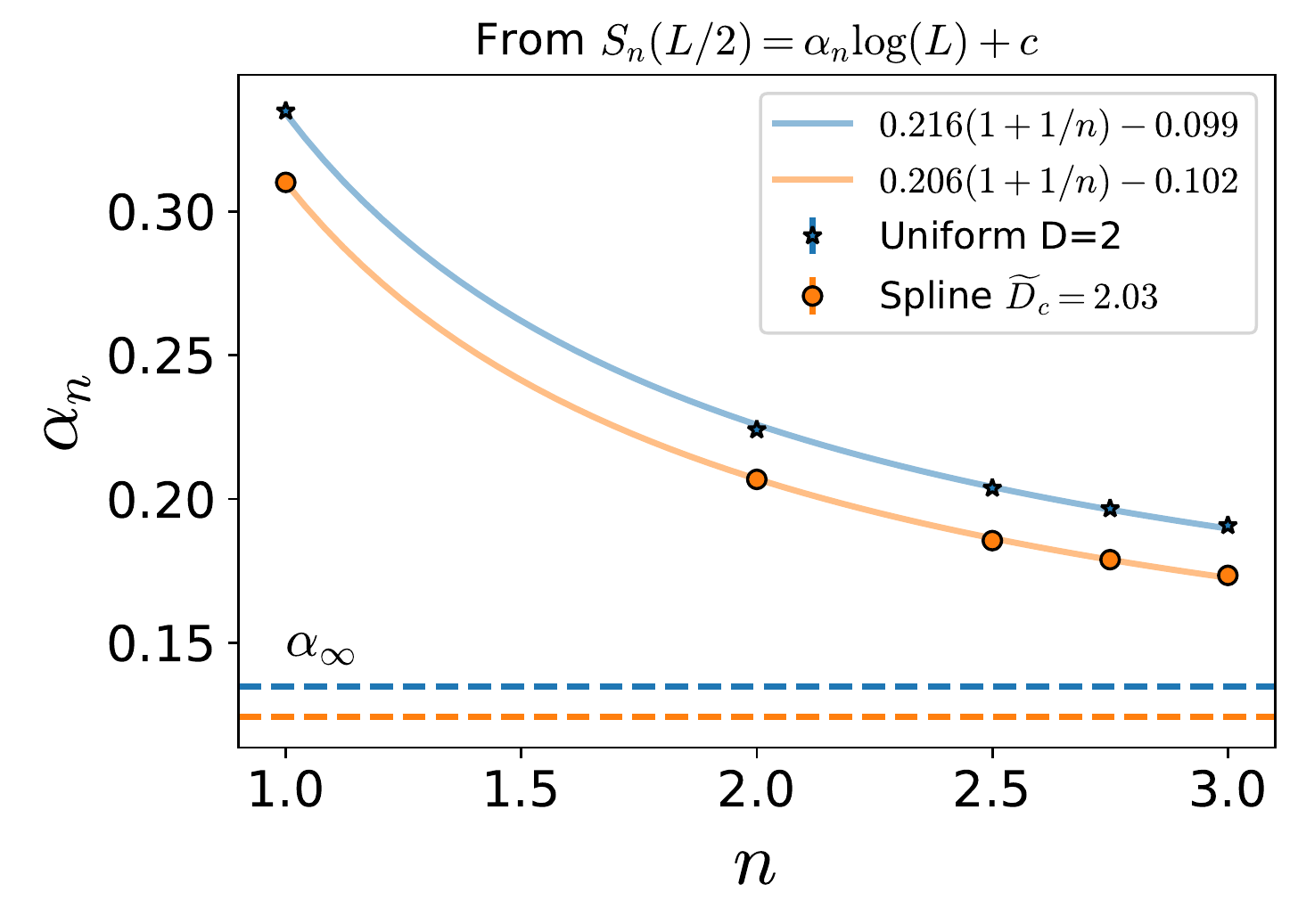}
    \caption{Behavior of the logarithmic coefficient at $D=2$ (stars) and $\dt=\dt_c=2.03$ (circles). To generate data at the critical point, we interpolate all available $\langle S_n(L/2) \rangle$ data with a 2D Bivariate Spline. The fit follows the ansatz from ref.~\cite{PhysRevB.101.060301}. We don't find exact agreement with disordered $\alpha_\infty\approx 0.124$  or uniform $\alpha_\infty \approx 0.135$ with a difference of about 0.02 for both.    }
    \label{fig:alpha_function}
\end{figure}

\subsection{Derivative Extrapolation}
\begin{figure}
    \centering
    \includegraphics[width=\columnwidth]{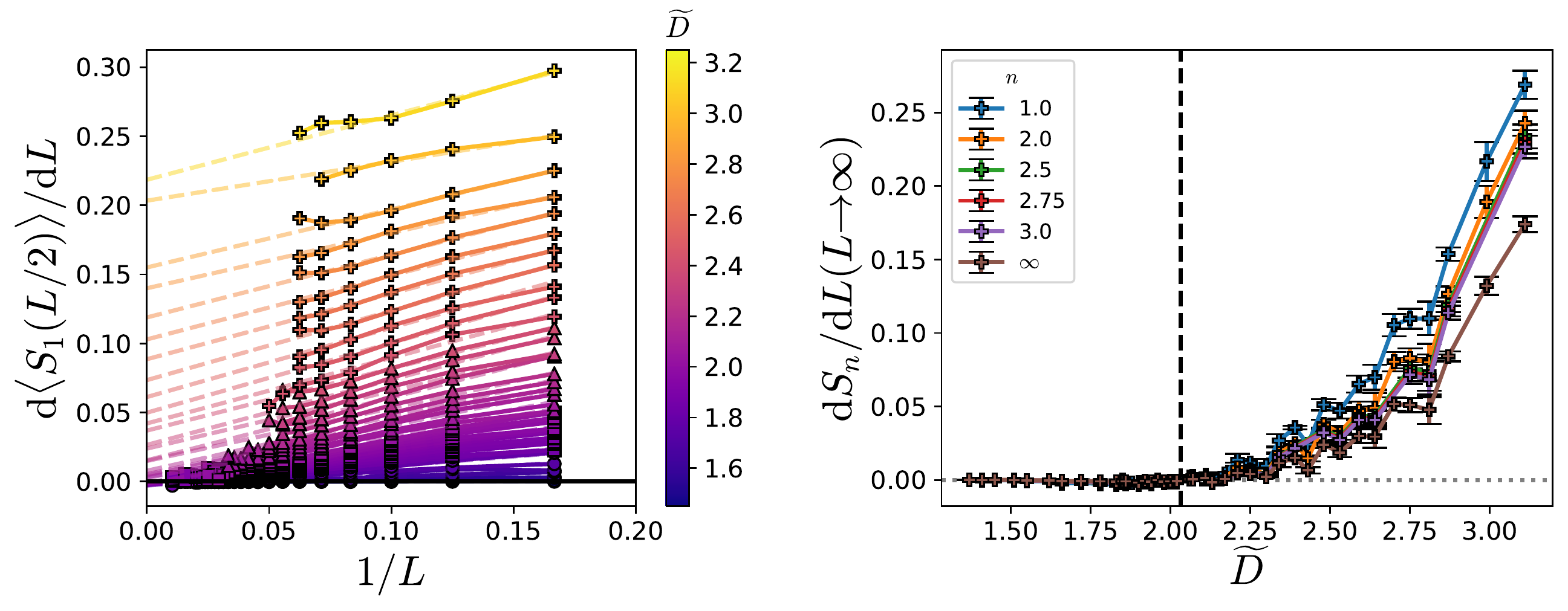}
    \caption{\textit{Left:} $d\langle S_1\rangle/dL$ derivative of the von Neumann entanglement entropy at the half cut as a function of $1/L$ using a spline fit. Shown in colored dashed lines is a linear fit of the derivative as a function of $1/L$ showing the infinite system limit value.
    \textit{Right:} Using the spline fits from $d\langle S_n\rangle /dL$ (left plot shows $n=1)$, we can obtain the estimated value at $1/L\to 0$ and the error from the fit. For $n=1$, we find a critical point of $\dt_c=2.033$ plotted as a vertical dashed line. }
    \label{fig:derivative}
\end{figure}

Using the averaged von Neumann entanglement entropy data we can try to observe the critical point data collapse. To do this, we use linear fits to the data, and compute the derivative at all known $L$ values. Taking a linear fit of the derivative vs $1/L$, we see in fig.~\ref{fig:derivative} the transition between near zero slope (or negative slope which is an error due to noise) to non-zero slope after $\dt \approx 2.03$, determined by fitting the infinite system values with a spline and finding its root. This value is consistent with the location of $\dt_c$ found from the tripartite mutual information.

\subsection{Higher R\'enyi Entropy Distributions}

\begin{figure}
    \centering
    \subfloat{\includegraphics[width=0.5\columnwidth]{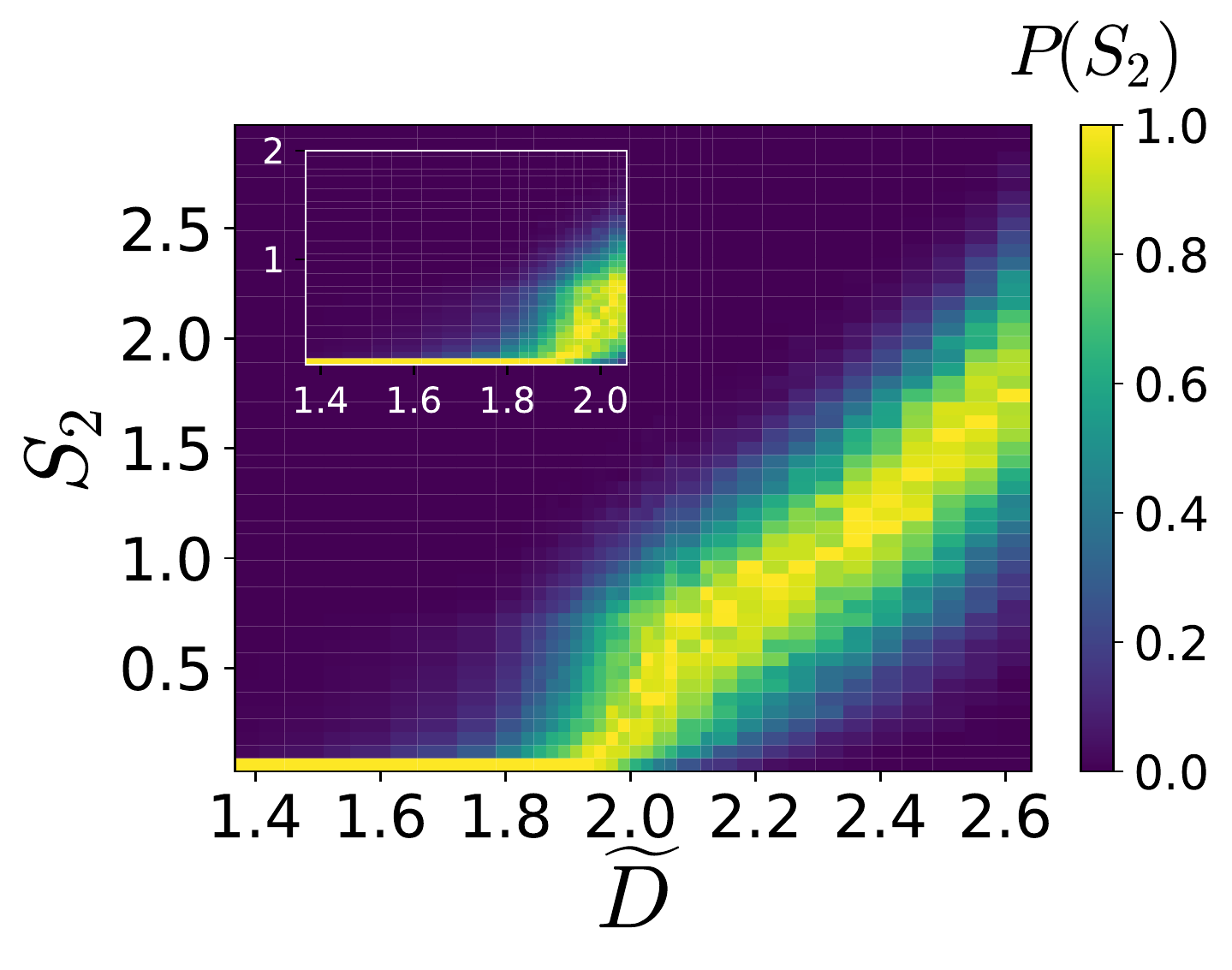}}
    \subfloat{\includegraphics[width=0.5\columnwidth]{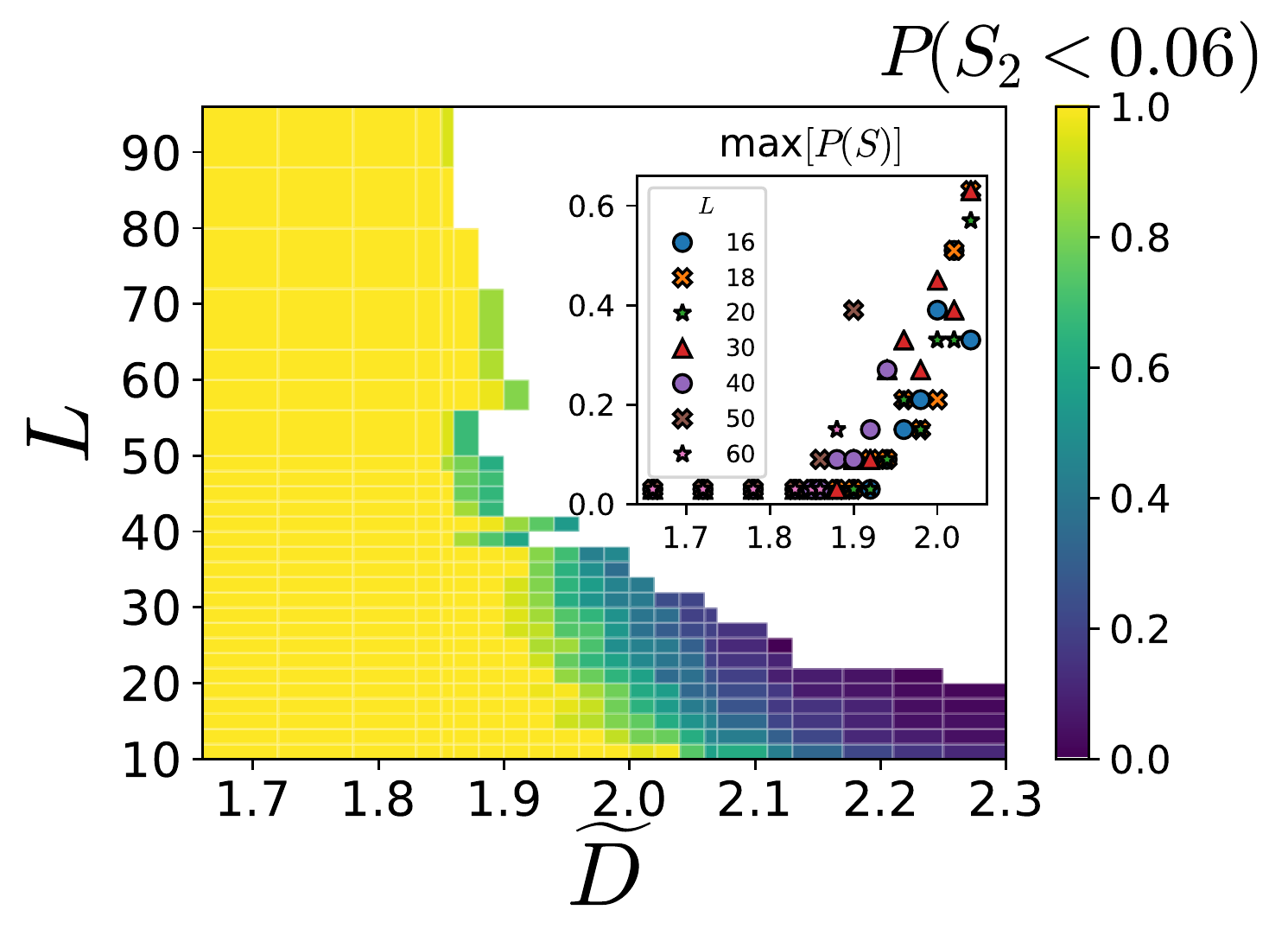}}
    
    \subfloat{\includegraphics[width=0.5\columnwidth]{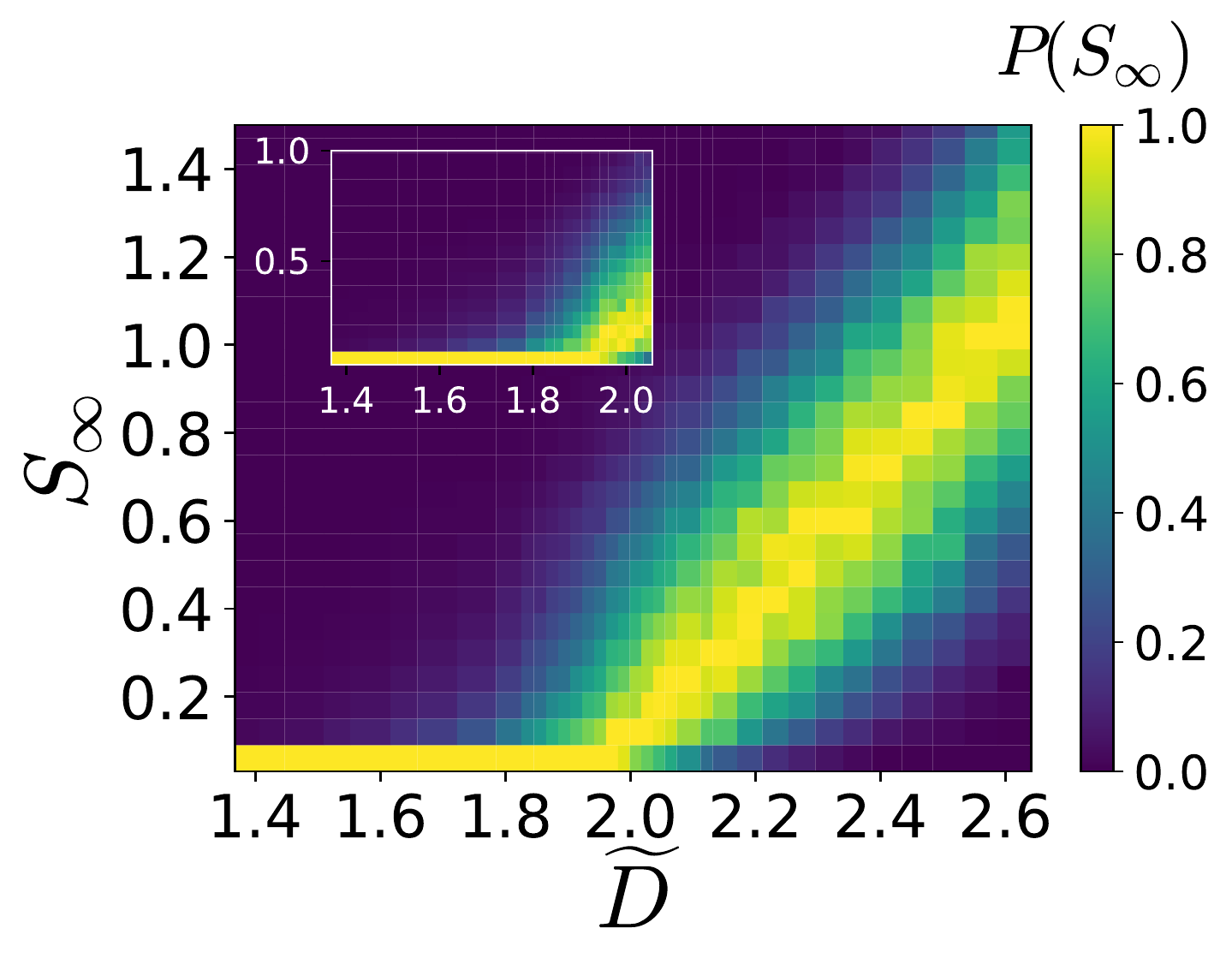}}
    \subfloat{\includegraphics[width=0.5\columnwidth]{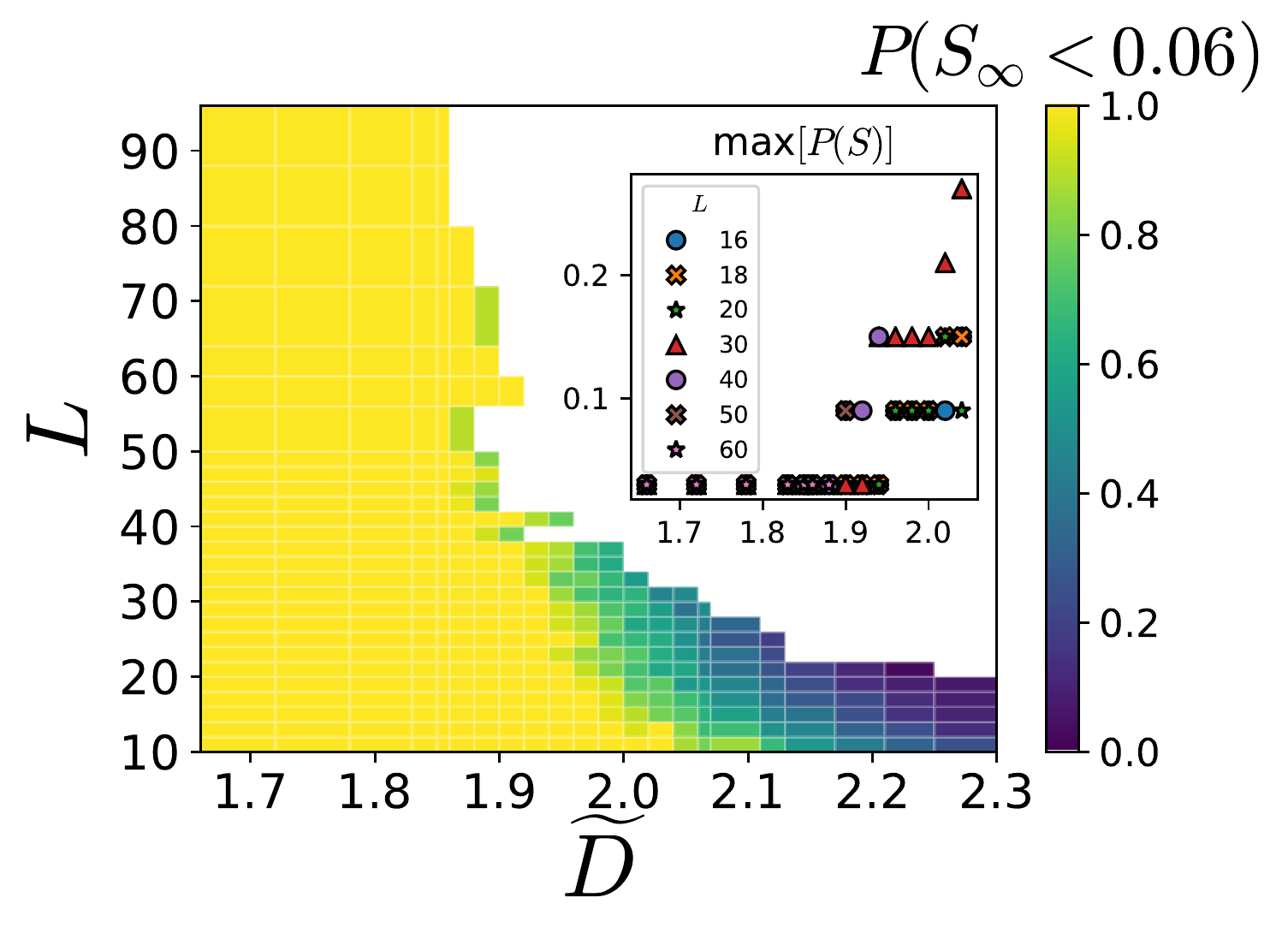}}
    
    \caption{
    \textit{Left:} normalized histogram of the R\'enyi entanglement entropy $S_n(L/2)$ for $n=2$ (top) and $n=\infty$ (bottom) as a function of $\dt$ for a system of width $L=16$ (\textit{inset:} $L=28$) with bins of width 0.06.
    \textit{Right:} system width $L$ dependence on $P(S_n<0.06)$ or nearly-zero entropy.  \textit{Inset:} mode of the binned entanglement entropy distribution as a function of $\dt$ for various $L$. 
    }
    \label{fig:S2inf_hists}
\end{figure}

We show the same data presented in fig.~3  for the $n=2$ and $n=\infty$ R\'enyi entropy in figs.~\ref{fig:S2inf_histograms}. Qualitatively the distributions show similar behavior for each R\'enyi entropy value. We can also compare figs.~4 and 5 for $n=2,\infty$ in fig.~\ref{fig:S2inf_histograms} and fig.~\ref{fig:S2inf_hists}, which shows change in the distribution mode as well as critical point shifting as a function of $n$. 

\subsection{Data Collapse Methodology}
Using the tripartite mutual information we obtain $\dt_c$ and $\nu$ using data collapse with a scaling ansatz of 
\begin{align}
    I_3(A:B:C) = f\left[(\dt -\dt_c)L^{1/\nu} \right]
\end{align}
where $f$ is a universal function. To measure goodness-of-fit for the collapse, we calculate the quality factor proposed in ref.~\cite{HoudayerFiniteScaling} 
\begin{align}
\label{eq:quality}
    Q = \frac{1}{\mathcal{N}} \sum_{ij} \frac{(y_{ij} - Y_{ij})^2}{dy^2_{ij}+dY^2_{ij}}
\end{align}
where $y_{ij},dy_{ij}$ are the observed entanglement entropy and error, and $Y_{ij},dY_{ij}$ are the estimates of the master curve value at the corresponding system size and $\dt$. This objective function and data are fed into SciPy's Nelder-Mead minimization function, which searches for the optimal $\dt_c$ and $\nu$. Given data in a range of $\dt$, we pick the lowest $Q$ value for 10 different initial $\dt_c,\nu$ values to assist in convergence. We repeat this process using 5 increasingly narrow ranges of $\dt$ data, using the 5 resultant $\dt_c,\nu$ to calculate the mean of the respective values. About this point we use the error bar method of ref.~\cite{PhysRevB.101.060301} and find a contour of value $1.3Q$ in $\rho,\nu$ space.
Note that we find the process to be relatively stable in $\dt_c$, which is reflected in the reported error in the main paper.

\subsection{Entanglement Entropy Data Collapse}

We perform data collapse on the entropy values directly, using the same method as the tripartite mutual information in the previous section. As there is a logarithmic term in the scaling ansatz given by ref.~\cite{Vasseur2018}, we use as our ansatz
\begin{align}
    \left|\langle S_n(L/2;\dt) \rangle - \langle S_n(L/2;\dt_c) \rangle\right| = f\left[(\dt -\dt_c)L^{1/\nu} \right]\label{eq:s1_ansatz}
\end{align}
for some universal function $f$. Here we iterate over a fixed $\dt_c$ in steps of 0.0015, optimizing for $\nu$, and choosing the $\dt_c$,$\nu$ pair that has the lowest quality $Q$. For each $\dt_c$, we use 4 increasingly narrow ranges of $\dt$ data, averaging the resultant $(\dt_c$,$\nu)$ pairs to calculate the mean. We report the lowest $\langle Q \rangle$ value, and use the same method as the tripartite QMI to produce the standard errors. 

\subsection{Higher R\'enyi Tripartite Mutual Information}

\begin{figure}
    \centering
    \includegraphics[width=\columnwidth]{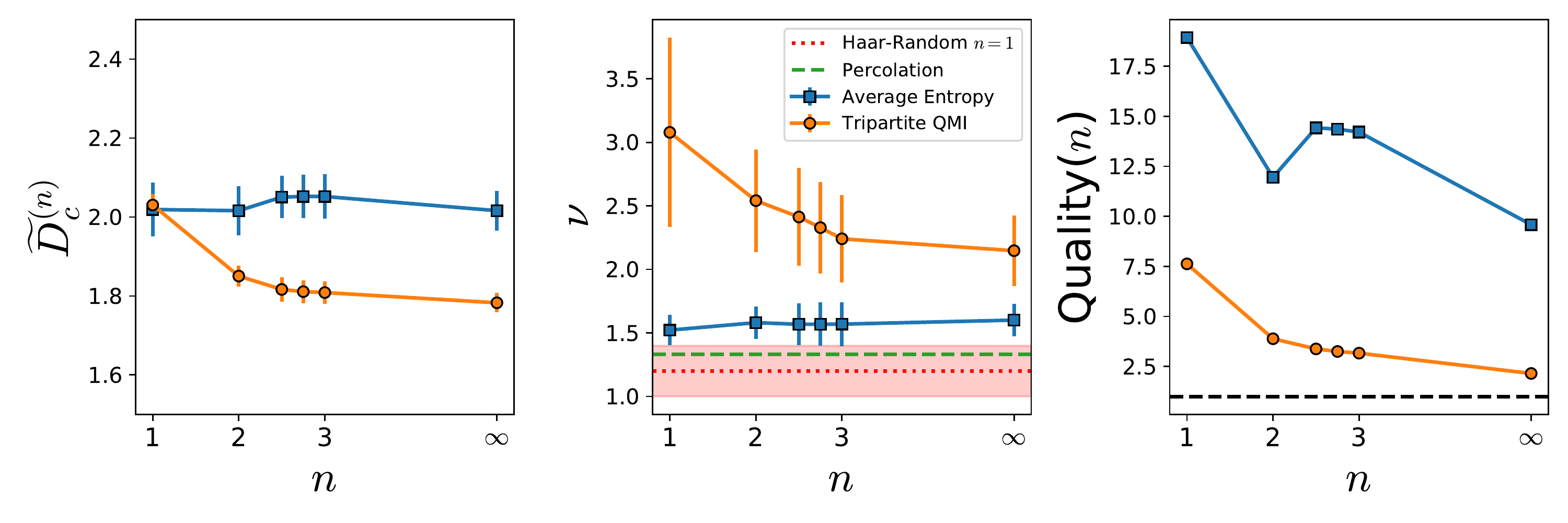}
    \caption{Results of performing data collapse using $I_3$ for the $n$th R\'enyi entropy (circles) and entanglement entropy $\langle S_n(L/2)\rangle $ (squares). The critical bond dimension (\textit{left}), exponent $\nu$ (\textit{middle}) and quality ($\textit{right}$, see eq.~\ref{eq:quality}) show a dependence on $n$ for the tripartite QMI. The value of $\nu$ for Haar-random quantum circuits taken from ref.~\cite{PhysRevB.101.060301} is denoted as a dashed red line with the broad red region representing reported error bars.}
    \label{fig:critical_exp}
\end{figure}
We perform data collapse of the tripartite mutual information and measure $\dt_c^{(n)}$, $\nu$, and the quality of the collapse in fig.~\ref{fig:critical_exp}. For all $n$, we find that $\nu > 1$ which indicates the bond dimension disorder remains irrelevant \cite{Vasseur2018}. The critical bond dimension has a weak dependence on $n$ which disagrees with the half-cut ($L_A=L/2$) R\'enyi entropy. The distribution qualitatively suggests the higher the index $n$, the higher the $\dt_c^{(n)}$, particularly above $\dt_c^{(1)}=2.02$. 

\subsection{Mutual Information Decay}
\begin{figure}
    \centering
    \includegraphics[width=\columnwidth]{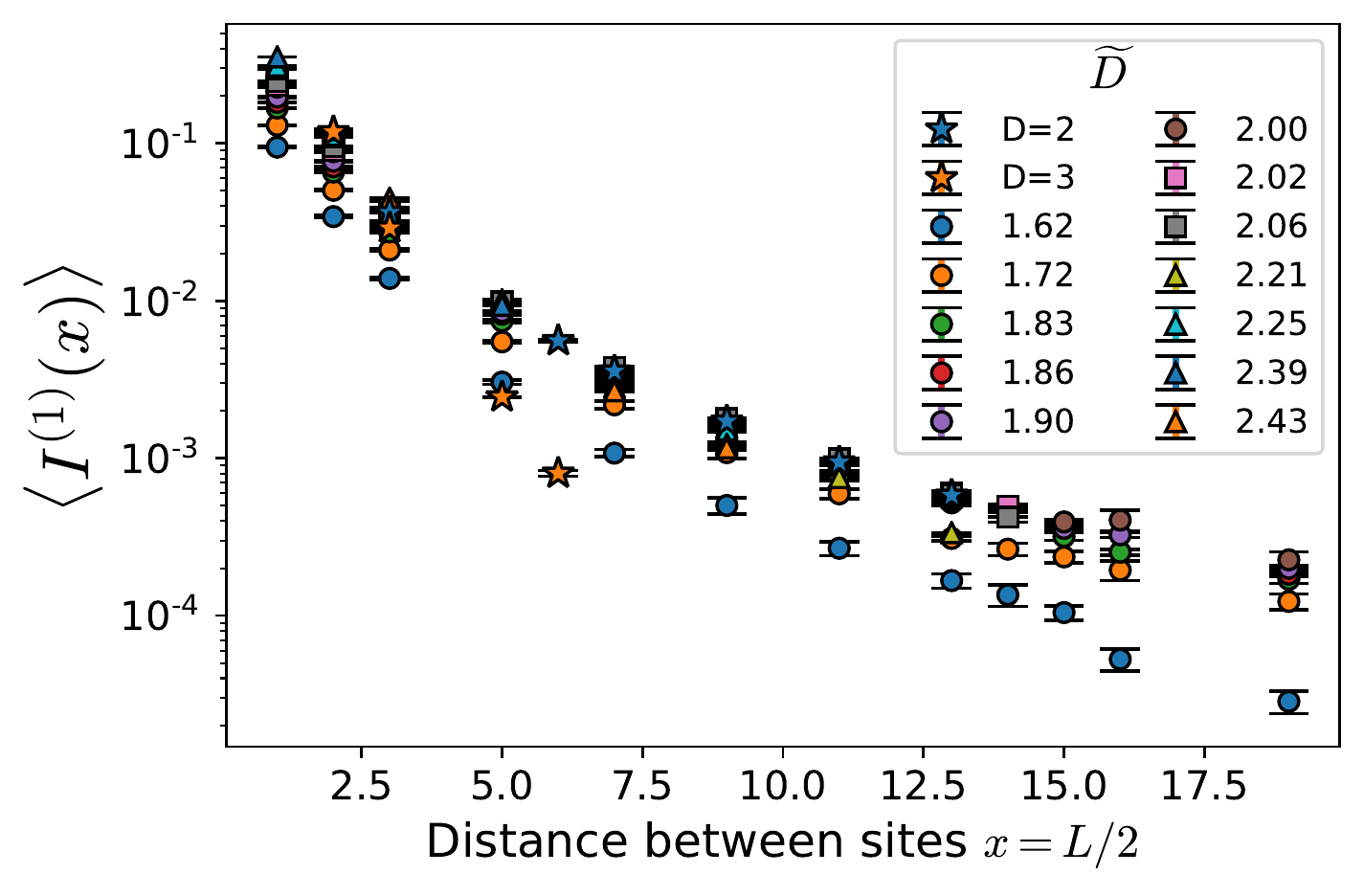}
    \caption{ Mutual information $\langle I^{(1)}(x)\rangle$ between two sites located a distance $x=L/2$ away for various values of $\dt$ on a semi-log plot to show exponential decay. $D=3$ and $\dt=1.67$ appears to show exponential decay, while most other data shows algebraic decay.
    }
    \label{fig:qmi_decay_semilog}
\end{figure}
\begin{figure}
    \centering
    \includegraphics[width=\columnwidth]{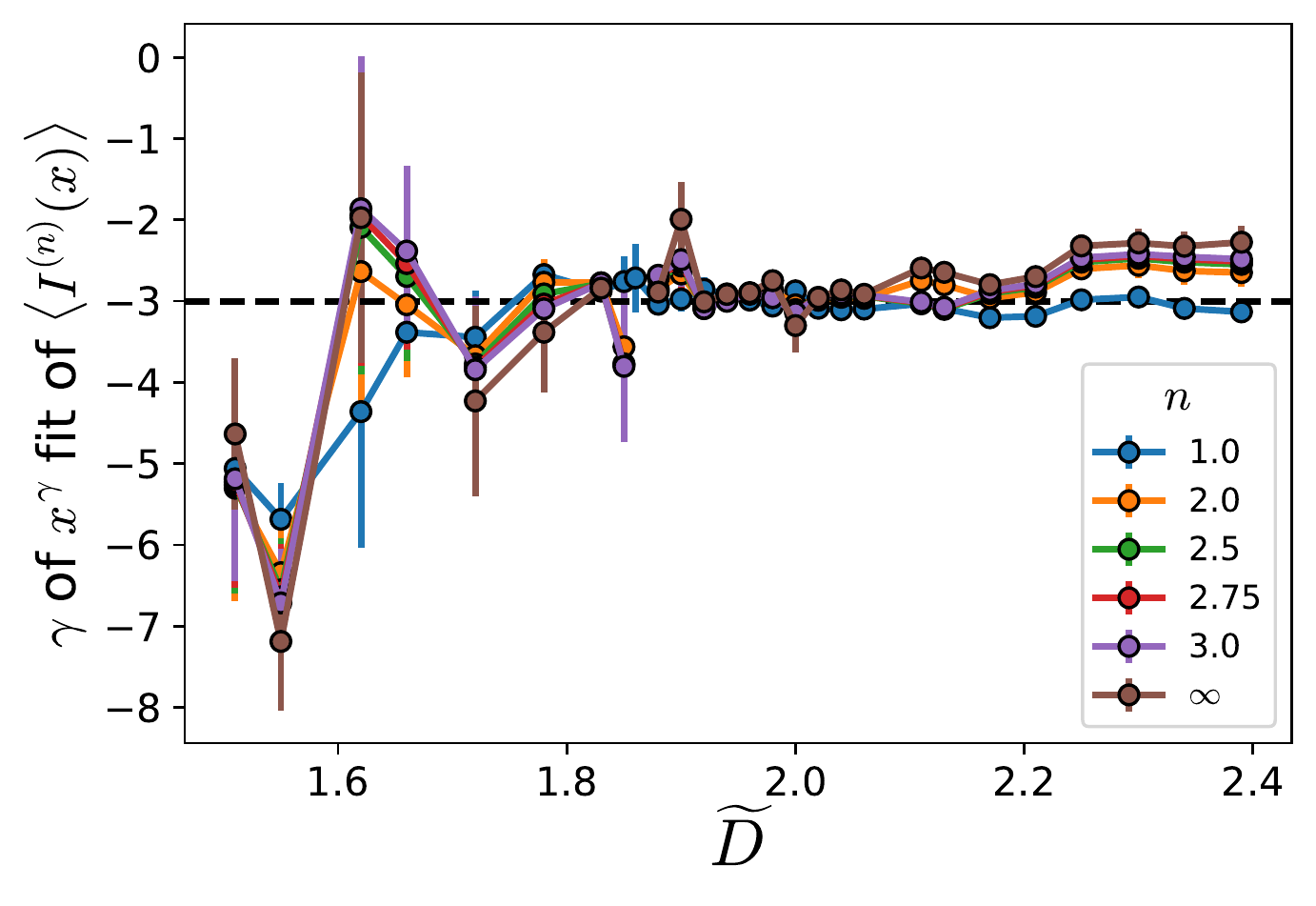}
    \caption{Fitting mutual information $\langle I^{(n)}(x)\rangle$ to a power-law decay $x^{\gamma}$, showing $\gamma$ vs $\dt$ for different R\'enyi indices. 
    }
    \label{fig:qmi_decay_slope}
\end{figure}

First we present fig.~\ref{fig:qmi_decay} as a semi-log plot in fig.~\ref{fig:qmi_decay_semilog}, as further evidence of the algebraic behavior far away from the critical point.   
Then, assuming we can fit the mutual entropy decay of two sites $\langle I^{(n)}(x) \rangle$ as a function of distance $x$ for different R\'enyi indices in fig.~\ref{fig:qmi_decay_slope}. There is little R\'enyi nor exponent dependence between $1.7\lesssim \dt \lesssim 2.4$, although we are unable to obtain data beyond $\dt \approx 2.4$. Note that this fit is very sensitive to noise in the data as well as the maximal value for distance $x$.

\begin{figure*}
    \centering
    \subfloat{\includegraphics[width=0.95\columnwidth]{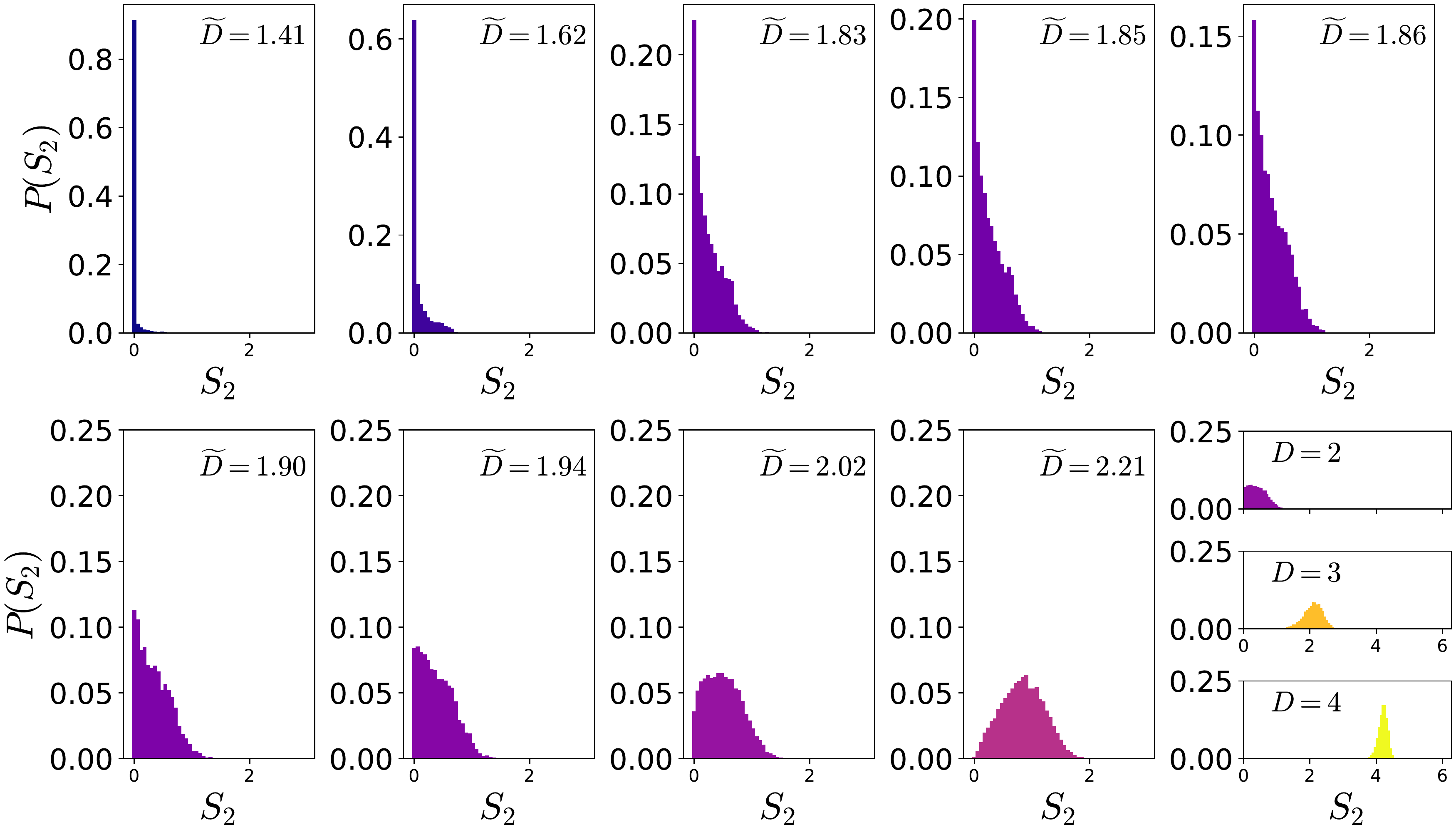}}
    \hspace{0.5cm}
    \subfloat{\includegraphics[width=0.95\columnwidth]{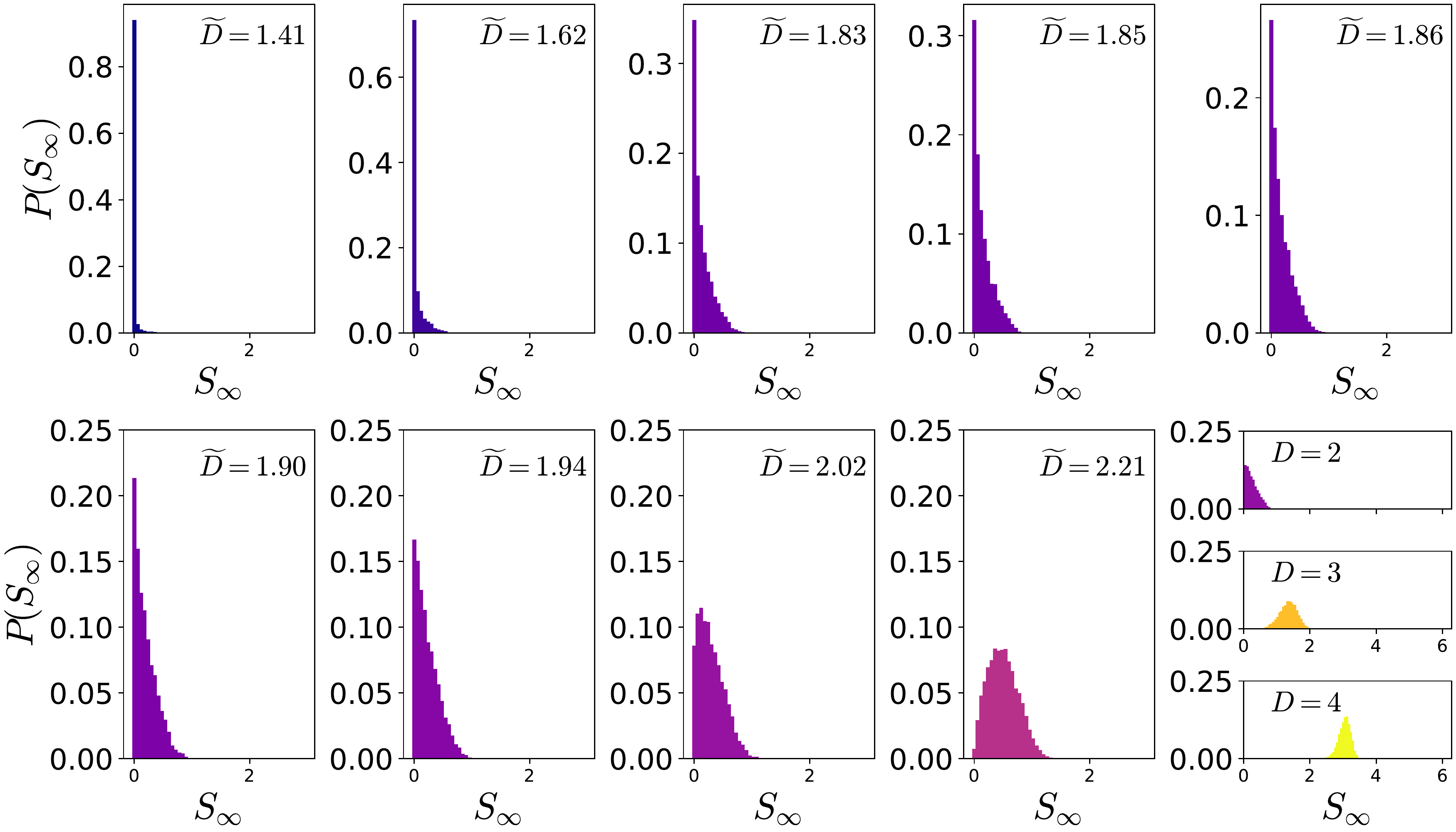}}
    \caption{Histogram of the $n=2$ (\textit{left}) and $n=\infty$ (\textit{right}) R\'enyi entanglement entropy $S_n(L/2)$ for various $\widetilde{D}$ for a system of width $L=18$. }
    \label{fig:S2inf_histograms}
\end{figure*}

\end{document}